\documentclass[aps,twocolumn,superscriptaddress,floatfix]{revtex4-1}
\usepackage{graphicx,amsfonts,amssymb,amsmath,mathrsfs,hyperref}

\newif\ifhyper
\hypertrue
\ifhyper
\hypersetup{
   citecolor = {red},
   colorlinks = {true}, 
   linkcolor = {blue},
   urlcolor = {blue} 
}
\fi

\def\be{\begin{equation}}
\def\ee{\end{equation}}
\def\bea{\begin{eqnarray}}
\def\eea{\end{eqnarray}}

\newcommand{\Zd}{\mathbb{Z}_2}

\newcommand{\Hi}{\mathcal{H}}

\newcommand{\C}{\mathbb{C}}

 \newcommand{\ket}[1]{|#1\rangle}
 \newcommand{\bra}[1]{\langle #1|}
 \newcommand{\braket}[2]{\langle #1|#2\rangle}
 \newcommand{\expectval}[1]{\langle #1 \rangle}

\begin{document}

\title{Infinite Projected Entangled-Pair State algorithm for ruby and triangle-honeycomb lattices}

\author{Saeed S. Jahromi}
\email{jahromi@physics.sharif.edu}
\affiliation{Department of Physics, Sharif University of Technology, Tehran 14588-89694, Iran}

\author{Rom\'an Or\'us}
\affiliation{Institute of Physics, Johannes Gutenberg University, 55099 Mainz, Germany}

\author{Mehdi Kargarian}
\affiliation{Department of Physics, Sharif University of Technology, Tehran 14588-89694, Iran}

\author{Abdollah Langari}
\affiliation{Department of Physics, Sharif University of Technology, Tehran 14588-89694, Iran}

\begin{abstract}

The infinite Projected Entangled-Pair State (iPEPS) algorithm is one of the most efficient techniques for studying the ground-state properties of two-dimensional quantum lattice Hamiltonians in the thermodynamic limit. Here, we show how the algorithm can be adapted to explore nearest-neighbor local Hamiltonians on the ruby and triangle-honeycomb lattices, using the Corner Transfer Matrix (CTM) renormalization group for 2D tensor network contraction. Additionally, we show how the CTM method can be used to calculate the ground state fidelity per lattice site and the boundary density operator and entanglement entropy (EE) on an infinite cylinder. As a benchmark,  we apply the iPEPS method to the ruby model with anisotropic interactions and explore the ground-state properties of the system. We further extract the phase diagram of the model in different regimes of the couplings by measuring two-point correlators, ground state fidelity and EE on an infinite cylinder. Our phase diagram is in agreement with previous studies of the model by exact diagonalization. 

\end{abstract}
\maketitle

%
%
\section{Introduction}

Studying different properties of quantum many-body systems and characterizing emergent phases of matter has been one of the biggest challenges of condensed matter physics. This has led to developments of many efficient numerical algorithms such as exact diagonalization (ED), quantum Monte Carlo \cite{Ceperley1980} and tensor network (TN) methods \cite{Verstraete2008,Orus2014}. TNs have already proved to provide an efficient representation for the ground-state of 1D gapped local Hamiltonians in the form of Matrix Product States (MPS) \cite{Fannes1992,Ostlund1995} which are the root of the well-known Density Matrix Renormalization Group (DMRG) method \cite{White1992,White1993}. The generalization of MPS to higher-dimensional systems has also been put forward by using Projected Entangled Pair States (PEPS) \cite{Verstraete2004,Verstraete2006}. Recent developments at both algorithmic and numerical levels have made the PEPS technique one of the most efficient and accurate numerical methods for capturing the ground-state properties of 2D quantum lattice models. The infinite version of the PEPS,  i.e., the infinite-PEPS (or iPEPS) \cite{Vidal2007,Corboz2010,Phien2015}, has also been developed to study 2D quantum lattice systems directly in the thermodynamic limit and has proven very successful in the study of the ground-state properties of many different models \cite{Orus2009,Corboz2012,Corboz2013,Matsuda2013,Corboz2014,Corboz2014a}. 

One of the problems that the iPEPS algorithm faces is the large computational cost of the contraction of the 2D infinite TN, and therefore approximation methods must be used. Different approaches have already been proposed for contraction of 2D TNs such as the boundary MPS \cite{Vidal2007}, tensor renormalization group (TRG) \cite{Levin2007, Gu2008}, and Corner Transfer Matrix (CTM) renormalization group \cite{Orus2009, Corboz2010a}, to name a few. In practice, each of these methods has its own benefits and problems. The recently developed iPEPS techniques \cite{Vidal2007,Corboz2010,Phien2015} based on CTMs have proven to be quite stable, accurate, fast and reliable. However, the downside is that the CTM method is only applied straightforwardly to the 2D square lattice, whereas for other lattice structures this may not be obvious at all. In any case, the method has also been successfully applied to other cases such as the honeycomb \cite{OsorioIregui2014} and kagome \cite{Corboz2012,Kshetrimayum2016a,Picot2016} lattices.

\begin{figure}[t]
\centerline{\includegraphics[width=\columnwidth,trim={1cm 1cm 0 1cm}]{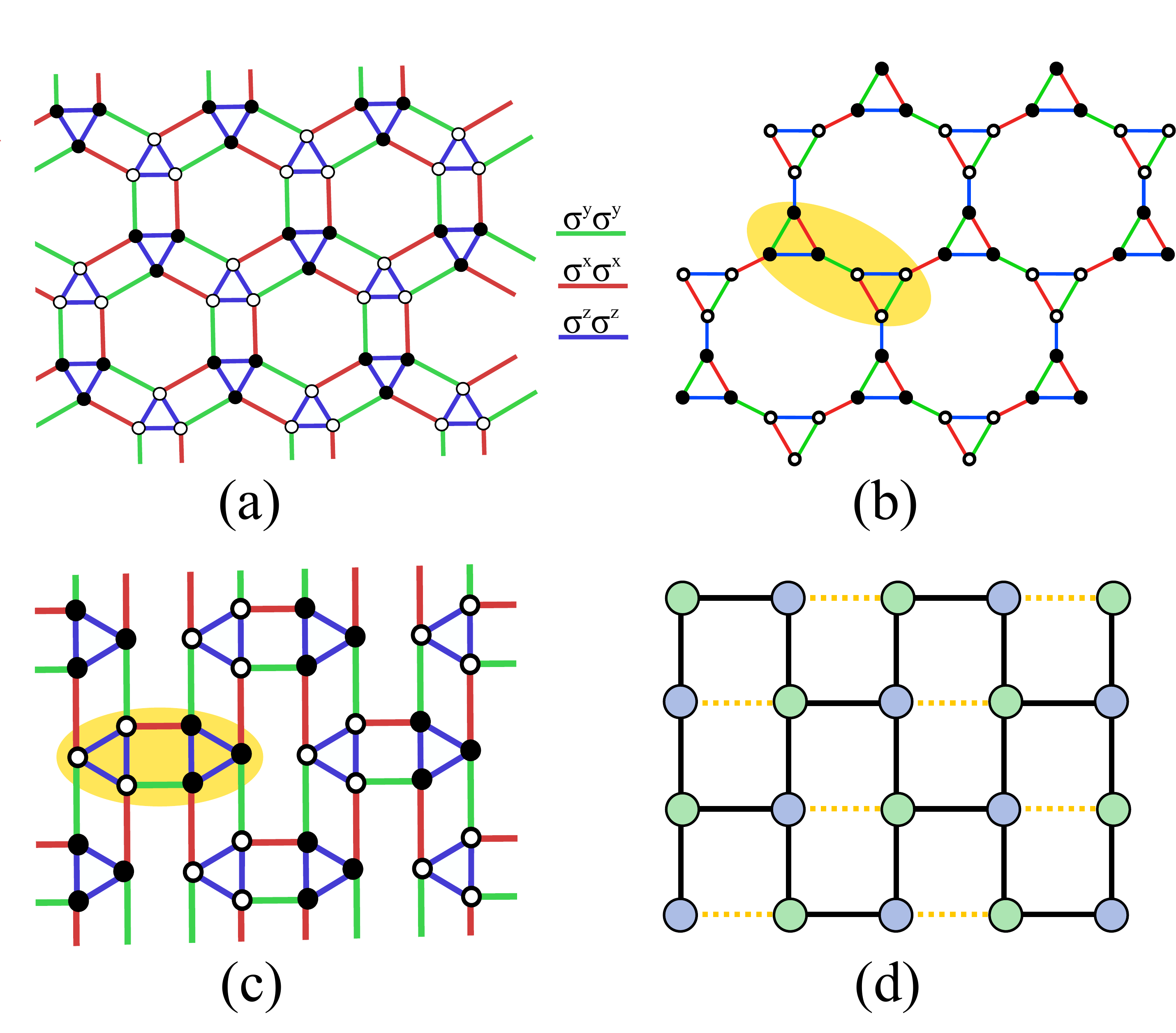}}
\caption{(Color online) (a) The ruby and (b) star lattices. Spin-$\frac{1}{2}$ particles are placed at the vertices of the lattices and the two-body interactions are characterized by different colors. (c) Ruby lattice reshaped into a brick-wall structure. Yellow region denotes a ruby unit cell with six spins, with sets of three spins placed at the vertices of a triangle and distinguished in black and white. (d) Each triangle in the brick-wall lattice (c) is replaced with a block-site with local Hilbert space dimension $2^3=8$. The resulting structure is topologically equivalent to a honeycomb lattice. The yellow dotted lines are auxiliary bonds with trivial interactions, which are added to the brick-wall lattice to form a square lattice.}
\label{Fig:ruby}
\end{figure}

In spite of the success of current TN algorithms for the study of quantum 
many-body systems, there are still many interesting models,
which are left behind due to their complicated interactions and lattice 
structures, thus making the  implementation challenging. The ruby 
(Fig.~\ref{Fig:ruby}-(a)) and star (Fig.~\ref{Fig:ruby}-(b)) lattices are two 
such examples, with interesting and rich underlying physics, especially 
concerning topologically-ordered (TO) phases. 

It has already been shown that anisotropic Kitaev interactions \cite{Kitaev2006} on the star lattice result in emergence of TO \cite{Tsui1982, Wen1995, Wen2007, Kitaev2003} chiral quantum spin liquids \cite{Yao2007, Kells2010, Dusuel2008}. Besides, anti-ferromagnetic Heisenberg interactions on the start lattice produce  ground states that lack magnetic order \cite{Richter2004,Farnell2011,Farnell2014,Yang2010}. The ruby model with anisotropic interactions \cite{Bombin2009,Kargarian2010}, Hamiltonian \eqref{eq:H-KR} (see also Fig.~\ref{Fig:ruby}-(a)), has also very interesting features such as hosting the topological color code (TCC) \cite{Bombin2006} (which is a quantum spin system aimed for the purpose of fault-tolerant quantum computation) as the low-energy effective theory of its gapped phase, supporting also string-net integrals of motion (IOM) \cite{Kargarian2010}. In contrast to the more conventional trivalent lattices with anisotropic Kitaev interactions, which are exactly solvable through Majorana fermionization \cite{Kitaev2006}, the ruby model cannot be solved exactly due to the four-valence structure of the ruby lattice, thus motivating its numerical study. Besides, the ruby lattice has physical realizations in bismuth ions of layered materials such as Bi$_{14}$Rh$_3$I$_9$, with interesting topological properties \cite{Rasche2013,Pauly2015,Pauly2016}. 

In this paper, we apply the iPEPS algorithm (based on CTMs) to the family of triangle-honeycomb structures such as ruby and star lattices. As a practical application, we study the ground state properties of the ruby model in the thermodynamic limit. In particular, we explore the low-energy properties and phase diagram of the system in different coupling regimes. We capture the quantum phase transitions of the ruby model by evaluating different quantities such as nearest-neighbor two-point correlators, entanglement entropy (EE) on an infinite cylinder \cite{Cirac2011, Schuch2013} and ground-state fidelity per lattice site \cite{Zhou2008}. Moreover, we present the details for the calculation of the ground state fidelity and EE on infinite cylinders using CTMs. 

This work is organized as follows: in Sec.~\ref{Sec:method} we briefly review the iPEPS technique and explain how to apply the method for the ruby lattice. In Sec.~\ref{Sec:Fidelity} we explain how to calculate the ground state fidelity per lattice site using CTMs. Details of the calculation of the EE on infinite cylinders by means of CTMs are given in Sec.~\ref{Sec:BoundaryRho}. Then, we apply the method to the ruby model introduced in Sec.~\ref{Sec:model} and discuss its ground state properties and zero-temperature phase diagram in the thermodynamic limit in Sec.~\ref{Sec:Phase-Diag}. Finally, we present our conclusions in Sec.~\ref{Sec:Conclude}.

\section{${\rm i}$PEPS basics}
\label{Sec:method}
In this section we briefly review the basic ideas behind the iPEPS technique and prescribe the details of the method for the family of triangle-honeycomb lattices. We specifically present the method for the ruby lattice. However, the extension to the star lattice is straightforward.

\subsection{Generalities}
Consider a 2D quantum lattice model with $N$ sites with local Hilbert space at each site described by $\C^d$. The full Hilbert space of the system is therefore 
given by $(\C^d)^{\otimes N}$, whose size grows exponentially with the size of the system. Thus, the problem of finding the relevant eigenstates of the  system is essentially intractable even for moderate system sizes. Luckily, it is sometimes possible to use PEPS tensors to store and represent some area-law states that approximate ground-states of 2D local Hamiltonians. As such, these states constitute a tiny, exponentially-small, but relevant corner of the Hilbert space, which can be parameterized efficiently. Generically, a 2D PEPS is given by 
\be
\ket{\Psi}=\sum_{\{s_{\vec{r}_i}\}_{i=1}^N}^d F \left( A_{s_{\vec{r}_1}}^{[\vec{r}_1]}, \ldots, A_{s_{\vec{r}_N}}^{[\vec{r}_N]} \right) \ket{s_{\vec{r}_1}, \ldots, s_{\vec{r}_N}},
\ee
where $\ket{s_{\vec{r}_i}}$ is the local basis of the site $i$ at position 
$\vec{r}_i$ according to the geometry of the 2D  lattice and 
$A_{s_{\vec{r}_i}}^{[\vec{r}_i]}$ are the local tensors. For the case of the 
square lattice, one has tensors of rank five at each site consisting of $dD^4$ 
complex coefficients, where $d$ is the {\it physical} dimension and $D$ is the 
{\it bond} dimension. Importantly, the bond dimension $D$ controls both the size 
of PEPS tensors and the maximum amount of entanglement that can be handled by
PEPS. The operation $F$ is a tensor trace that contracts the bond indices 
of the tensors $A_{s_{\vec{r}_i}}^{[\vec{r}_i]}$.

In order to approximate the ground state of a given quantum lattice Hamiltonian, one can evolve the system in imaginary time $\beta$ (similar to the Time-Evolving Block Decimation (TEBD) method in 1D \cite{Vidal2007a,Orus2008}) i.e,
\be
\ket{\Psi_{\rm GS}}=\underset{\beta \to \infty}{\lim} \frac{e^{-\beta H} \ket{\Psi_0}}{||e^{-\beta H} \ket{\Psi_0}||}, 
\ee
with $\ket{\Psi_0}$ some appropriate initial state. Efficient numerical algorithms have already been developed for both finite \cite{Verstraete2004,Verstraete2006} and infinite PEPS \cite{Vidal2007,Corboz2010,Phien2015} based on imaginary time evolution of translationally invariant local Hamiltonians on the square lattice. In particular, recent versions of the iPEPS method use the so-called {\it fast full update} \cite{Phien2015} for a stable and fast updating procedure of the tensors. Moreover, it has become clear that methods based on CTMs are particularly well-suited to approximate effective environments and estimate expectation values of local observables for infinite 2D lattices \cite{Orus2009}. 

In the next subsection, we show how to map the ruby lattice to a brick-wall structure (the procedure for the star lattice and other Archimedean lattices \cite{Richter2004a} is similar. See also Appendix.~\ref{appx:star-block} for more details on the iPEPS implementation of the star lattice), so that the iPEPS method for the square lattice \cite{Orus2009,Phien2015} is also applicable for the family of triangle-honeycomb lattices.

\subsection{Ruby lattice and trotterization}
 
 \begin{figure}[t]
\centerline{\includegraphics[width=\columnwidth]{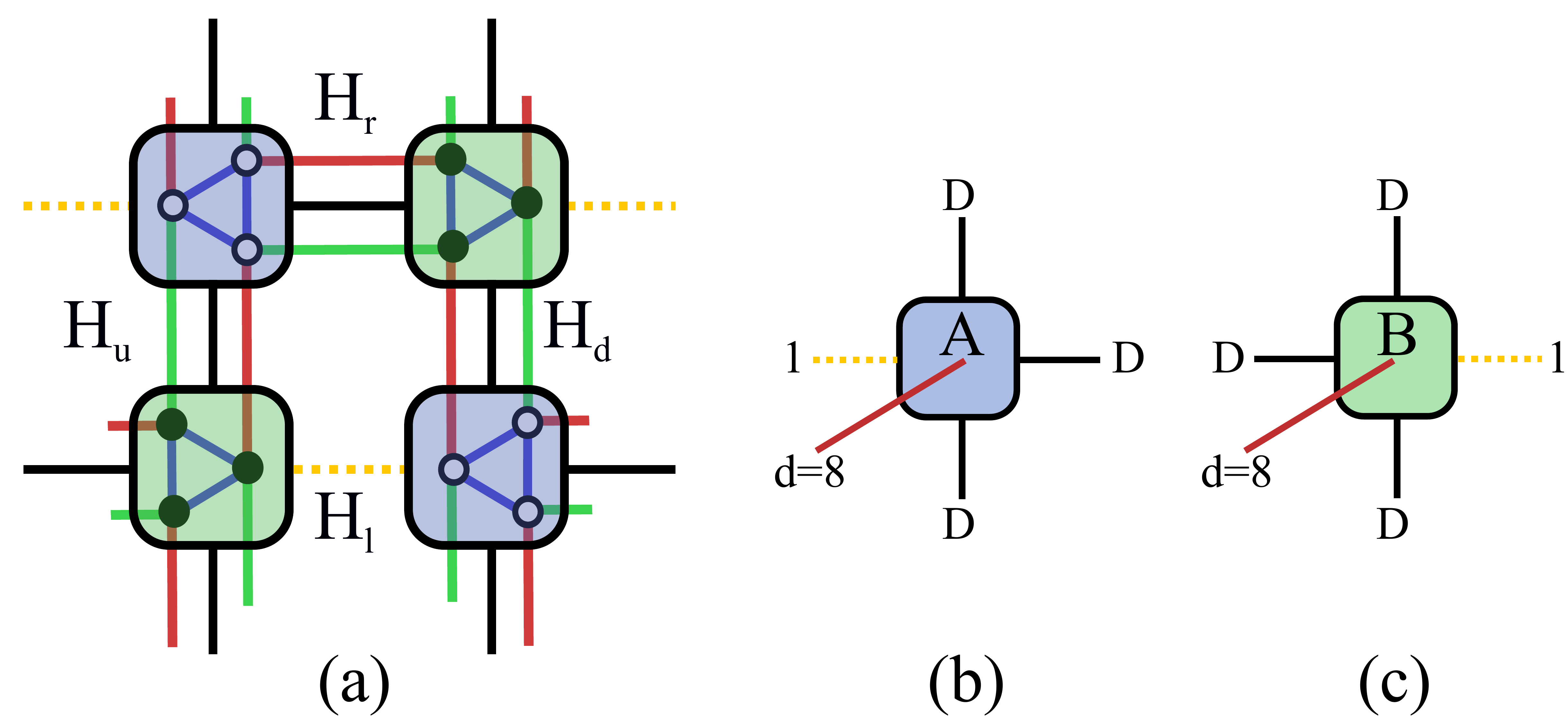}}
\caption{(Color online) (a) Tensor network structure of a $2\times 2$ unit cell of the square lattice of block sites. (b), (c) $A$, $B$ iPEPS tensors for block sites in a ruby unit cell. Each tensor has a physical index with dimension $2^3$ which describes the local Hilbert space of the three spins in each block site, three bond indices with dimension $D$ each  (accounting for the directions with interactions), and one trivial bond index with dimension $1$.}
\label{Fig:unitcel}
\end{figure}

Let us now consider how to adapt the iPEPS methodology to the case of the ruby lattice.  Fig.~\ref{Fig:ruby}-(c) illustrates how the ruby lattice can be shaped into a brick-wall structure, which in turn is topologically equivalent to a honeycomb lattice of coarse-grained sites. Each unit cell of the ruby lattice is composed of six physical degrees of freedom (yellow region in Fig.~\ref{Fig:ruby}-(c)). Replacing each triangle in the unit cell with an effective block site with local Hilbert space of dimension $d=2^3$, we end up with a brick-wall honeycomb lattice (see Fig.~\ref{Fig:ruby}-(d)). Next, by associating an iPEPS tensor to each block site and introducing trivial indices \cite{OsorioIregui2014} as in the yellow dotted lines in Fig.~\ref{Fig:ruby}-(d), we end up with an iPEPS on the square lattice and a $2 \times 2$ unit cell specified by two tensors $A$ and $B$ according to a checkerboard pattern, see Fig.~\ref{Fig:unitcel}.

In order to approximate the ground-state of the system by imaginary time evolution, we consider the Hamiltonian of the system to be composed of only translationally invariant local terms with nearest-neighbor interactions i.e,
\be
H=\sum_{\langle \vec{r},\vec{r'} \rangle} h^{[\vec{r},\vec{r}']},
\ee
where, the sum runs over the nearest-neighbors $\vec{r}$ and $\vec{r}'$. Next, we approximate the imaginary time evolution operator in terms of two-body gates. To this end, we first write the Hamiltonian of the system as the sum of four mutually-commuting terms, i.e.,
\be
H=H_r+H_l+H_u+H_d,
\ee
where, $(r,l,u,d)$ denote the (right, left, up, down) links shown in Fig.~\ref{Fig:unitcel}-(a). For the ruby model, the explicit form of the $H_i$, $i\in$ $(r,l,u,d)$, are provided in Appendix.~\ref{appx:KR-IPEPS} (see also Ref.~\cite{Corboz2012a}). 

\begin{figure} [t]
\centerline{\includegraphics[width=\columnwidth]{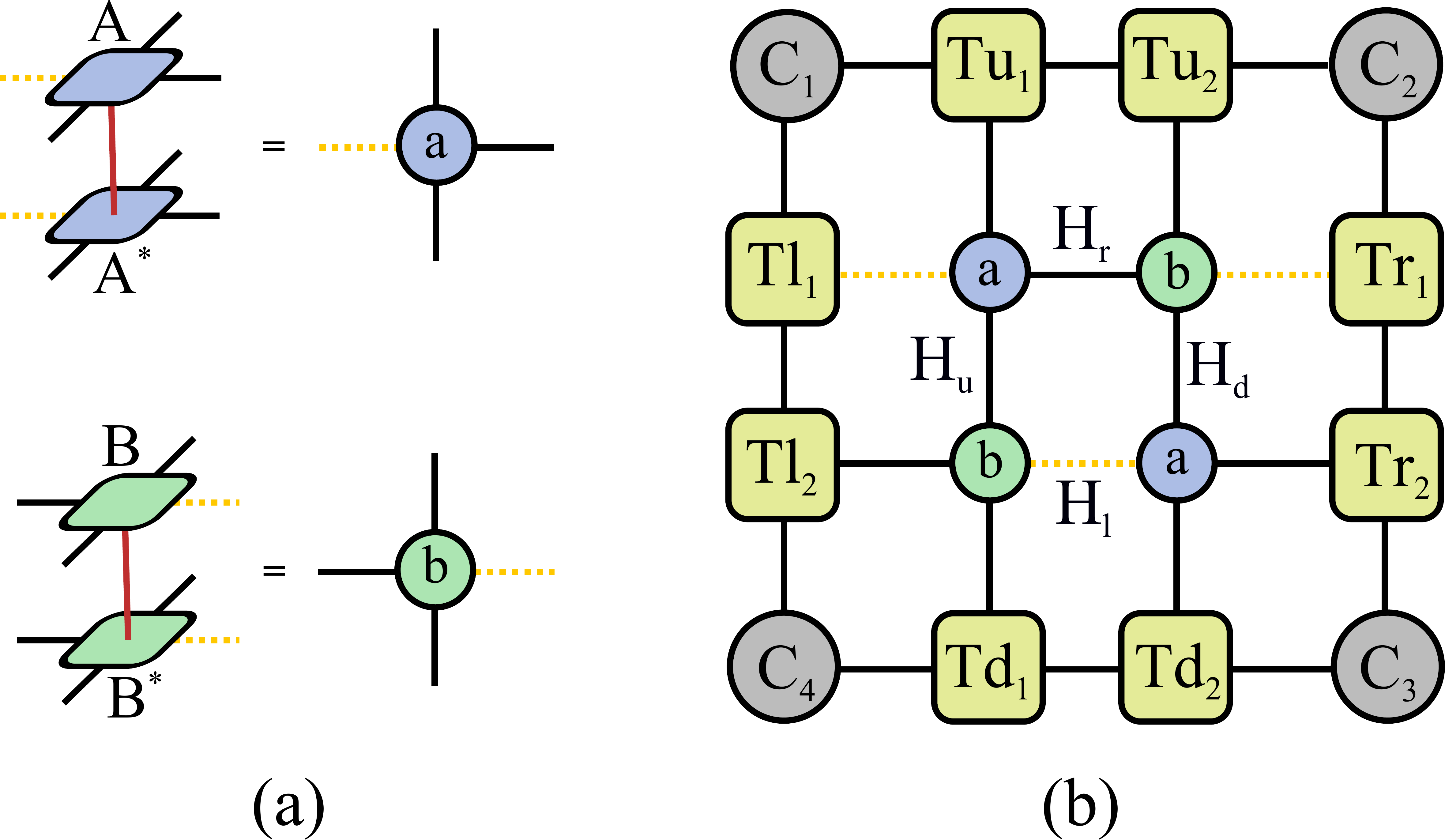}}
\caption{(Color online) (a) Reduced tensor $a$ ($b$) constructed from the contraction of the $A$ and $A^*$ ($B$ and $B^*$) tensors along their physical index. (b) Approximate contraction of an infinite 2D lattice in terms of an effective environment and a $2 \times 2$ unit cell of reduced tensors. The effective environment is composed by four $\chi\times\chi$ corner transfer matrices $\{C_1,C_2,C_3,C_4\}$ and eight half-column/half-row  $\chi\times\chi\times D^2$ transfer matrices  $\{T_{u_1},T_{u_2},T_{r_1},T_{r_2},T_{d_1},T_{d_2},T_{l_1},T_{l_2}\}$. See Ref.~\cite{Orus2009} for more details. }
\label{Fig:ctm}
\end{figure}
We then decompose the time evolution operator in terms of infinitesimal time steps $\delta \tau \equiv \beta/m \ll 1$, by applying a {\it Suzuki-Trotter} decomposition
\bea
e^{-\beta H}&=&(e^{-\delta\tau H})^m\\
&\approx&(e^{-\delta\tau H_r} e^{-\delta\tau H_l} e^{-\delta\tau H_u} e^{-\delta\tau H_d})^m.
\eea
In the above expression we applied a first-order decomposition, but higher orders are also possible and sometimes also convenient. Since each term $H_i$, is a sum of mutually commuting terms \cite{Phien2015}, we can further write $e^{-\delta\tau H_i}$ exactly as a product of two-body gates, i.e.,
\be
e^{-\delta\tau H_i} =\prod_{\langle \vec{r},\vec{r}'\rangle \in i} g^{[\vec{r},\vec{r}']}
\ee
where $g^{[\vec{r},\vec{r}']}\equiv e^{-\delta\tau h^{[\vec{r},\vec{r}']}}$.

The algorithm proceeds by applying these gates sequentially on every type of link, and replacing their effect over the whole lattice by translation invariance. In practice, we use the full update technique combined with a gauge fixing  \cite{Phien2015} to evaluate the effect of such gates in a $2\times 2$ unit cell. This procedure is repeated iteratively until some convergence criterion is fulfilled.

\subsection{Effective environments with CTMs}

Although PEPS are a very efficient way of representing approximations to relevant eigenstates of local Hamiltonians, the calculation of expectation values, and even scalar products between PEPS, is quite challenging, since the contraction of a 2D TN is in general an $\sharp$P-hard problem \cite{Schuch2007} and therefore approximations need to be used.  Here, we use the directional CTM method introduced in Ref.~\cite{Orus2009}, as well as a refined version of it  \cite{Corboz2010a} in order to approximate the environment of the iPEPS tensors around a $2\times 2$ unit cell. This is important in several scenarios, namely: for the calculation of expectation values, in the full update procedure, and also in the calculation of the ground-state fidelity per lattice site  \cite{Zhou2008} (see Sec.~\ref{Sec:Fidelity}) and the EE on a cylinder \cite{Cirac2011,Schuch2013} (see Sec.~\ref{Sec:BoundaryRho}).

The effective environment \cite{Orus2009}, also in our implementation of the ruby lattice, is given
in terms of four $\chi\times\chi$ corner transfer matrices $\{C_1,C_2,C_3,C_4\}$ and eight half-column/half-row  $\chi\times\chi\times D^2$ transfer matrices  $\{T_{u_1},T_{u_2},T_{r_1},T_{r_2},T_{d_1},T_{d_2},T_{l_1},T_{l_2}\}$ which surround a $2 \times 2$ unit cell (see Fig.~\ref{Fig:ctm}). The accuracy of these tensors is further controlled by the bond dimension $\chi$ of the environment. 

\begin{figure}
	\centering
	\includegraphics[width=\linewidth]{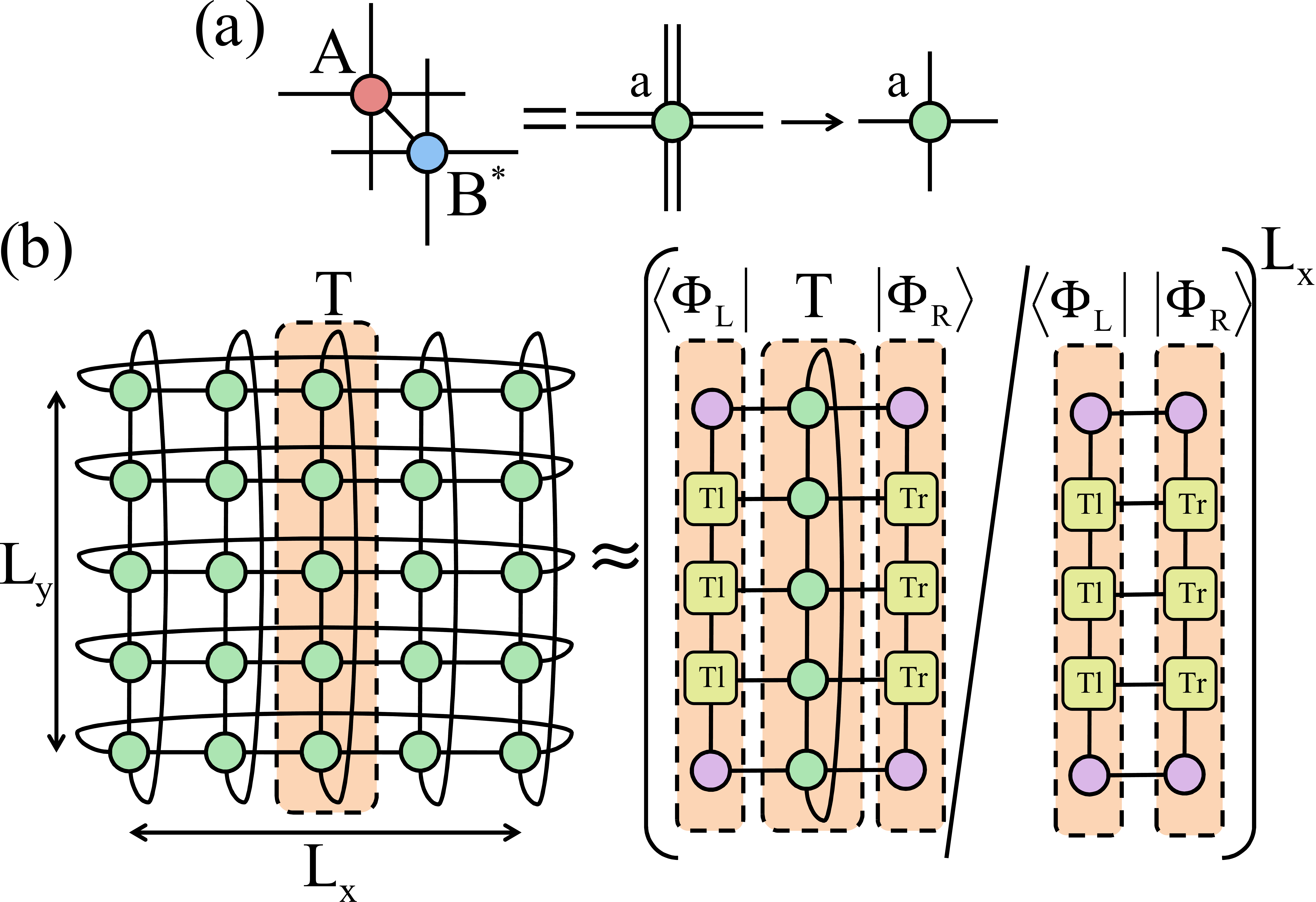}
	\caption{(Color online) (a) Reduced tensor to compute the fidelity. $A$ is the PEPS tensor for $\ket{\psi(\lambda_1)}$, $B$ is the PEPS tensor for $\ket{\psi(\lambda_2)}$. (b) Contraction leading to Eq.~(\ref{fi}).}
	\label{fig1}
\end{figure}

\subsection{Implementation details}
In order to approximate the ground-state of the systems in this paper, we use the full update approach based on imaginary time evolution with $\delta\tau=0.01$, accompanied by a proper choice of gauge fixing in the algorithm according to Ref.~\cite{Phien2015,Lubasch2014}. In order to accelerate the simulations further, we first approximate the ground-state tensors by applying a simple-update \cite{Jiang2008, Corboz2010} and then refine the iPEPS tensors by using the full update (with gauge fixing) and taking care of possible local minima. This helps to improve both the stability and the convergence of the algorithm. In our simulations we went up to bond dimensions $D=12$ for the iPEPS and $\chi = 100$ for the environment.

\section{Ground state fidelity from CTMs}
\label{Sec:Fidelity}

\begin{figure}
	\centering
	\includegraphics[width=\linewidth]{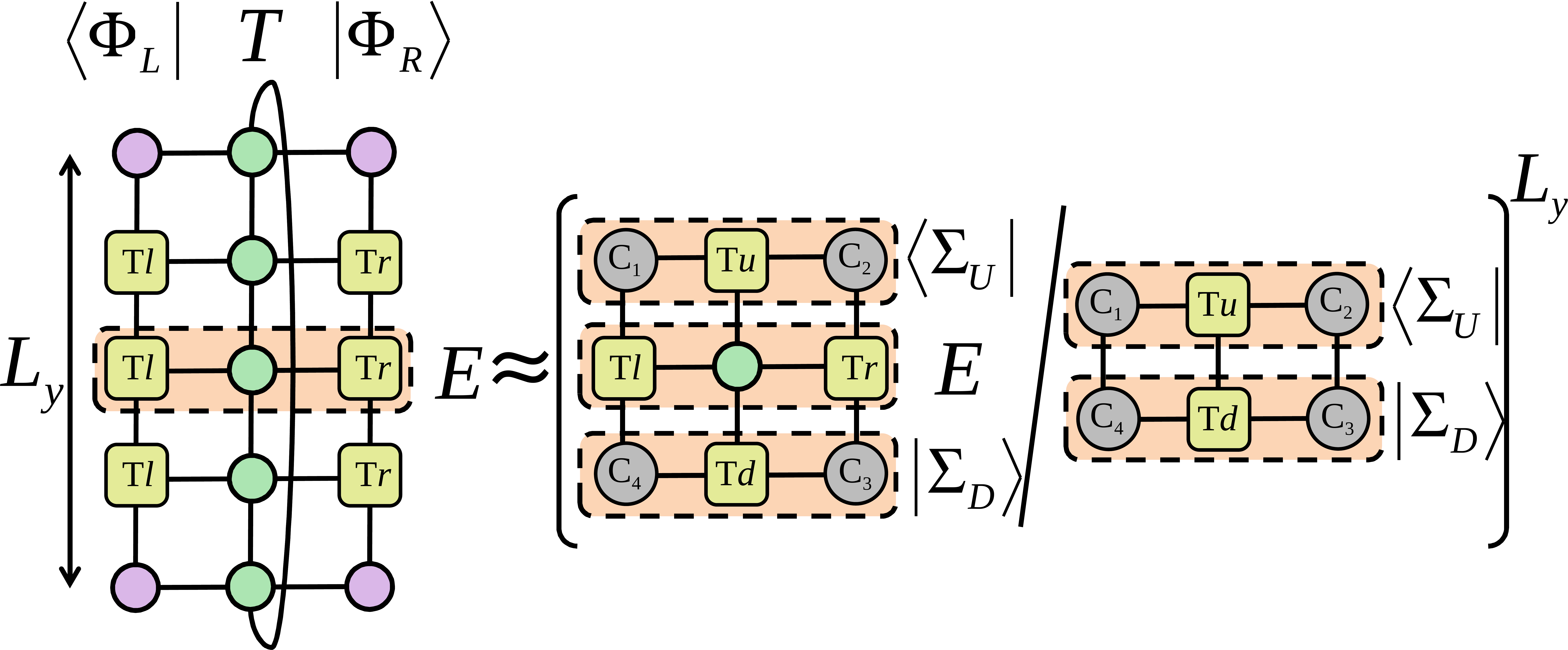}
	\caption{(Color online) Contraction leading to Eq.~(\ref{eq7}).}
	\label{fig2}
\end{figure} 

In this section, we explain how one can compute the fidelity per lattice site using CTMs, much more efficiently and accurately than using boundary MPS. Let us start by reminding ourselves of few basic concepts about the fidelity approach (see, e.g., Ref.~\cite{Zhou2008}). Consider a quantum lattice system with Hamiltonian $H(\lambda)$, $\lambda$ being a control parameter. For two different values $\lambda_1$ and $\lambda_2$ of this control parameter, we have ground-states $\ket{\Psi (\lambda_1)}$ and $\ket{\Psi(\lambda_2)}$. The ground-state fidelity is then given by $F(\lambda_1, \lambda_2) = |\braket{\Psi(\lambda_2)}{\Psi(\lambda_1)}|$, which scales as $F(\lambda_1, \lambda_2) \sim d(\lambda_1, \lambda_2)^N$, with $N$ the number of lattice sites. One therefore defines the \emph{fidelity per lattice site} as
\be
\ln d(\lambda_1, \lambda_2) \equiv \lim_{N \rightarrow \infty} \frac{\ln F(\lambda_1, \lambda_2)}{N}. 
\ee

Let us now explain how one can compute the above quantity very efficiently using the CTM formalism. First, as was noticed in Ref.~\cite{Zhou2008}, the fidelity for two ground-states represented by 2D PEPS can actually be mapped to the contraction of a 2D TN, amounting to the calculation of a classical partition function (the fidelity per site being the analog of a free energy). Let us assume, for the sake of simplicity, that both PEPS have a $1$-site unit cell. Then we have that 
\be
F(\lambda_1, \lambda_2) = \left|Ê{\rm tr} \left( T^{L_x} \right) \right| , 
\ee
with $L_x$ the horizontal number of sites of the PEPS, and $T$ the 1D transfer matrix  shown in Fig.~\ref{fig1}-(b). For $L_x \gg 1$, one has that 
\be
F(\lambda_1, \lambda_2) \sim \left|Ê\mu_0^{L_x} \right|Ê
\ee
with $\mu_0$ the dominant eigenvalue of transfer matrix $T$. In terms of the dominant left- and right-eigenvectors of $T$, this means that 
\be
F(\lambda_1, \lambda_2) \sim \left|Ê \frac{\bra{\Phi_L} T \ket{\Phi_R}}{\braket{\Phi_L}{\Phi_R}} \right|^{L_x} , 
\label{fi}
\ee
with $\ket{\Phi_L}$ and $\ket{\Phi_R}$ the dominant left and right eigenvectors of $T$ respectively, which we assume to be not necessarily normalized (hence the denominator in the above equation). The expression in Eq.~(\ref{fi}) corresponds to the tensor network diagram in Fig.~\ref{fig1}-(b).

\begin{figure}
	\centering
	\includegraphics[width=0.9\linewidth]{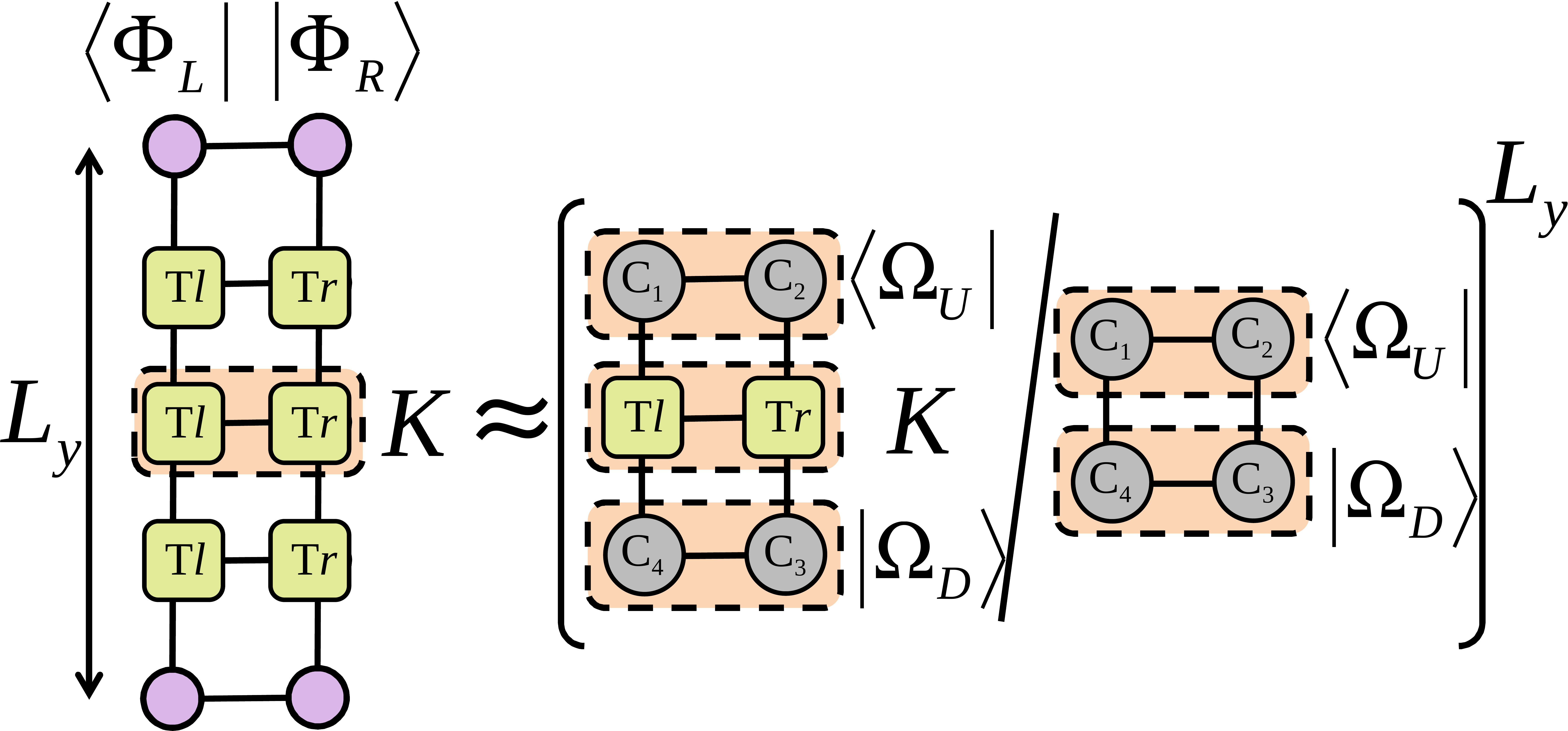}
	\caption{(Color online) Contraction leading to Eq.~(\ref{eq10}).}
	\label{fig3}
\end{figure}

Let us now focus on the numerator of Eq.~(\ref{fi}). Forgetting about the $L_x$ exponent, the term $\bra{\Phi_L} T \ket{\Phi_R}$ can be understood, as shown in Fig.~\ref{fig1}-(b), as the overlap between two MPS and a matrix product operator MPO for the 1D transfer matrix $T$. Thus we have the equation 
\be
\bra{\Phi_L} T \ket{\Phi_R}  = {\rm tr}Ê\left( E^{L_y} \right), 
\ee
with $E$ the 0D transfer matrix in Fig.~\ref{fig2}. Thus, in the limit $L_y \gg 1$ we have 
\be
\bra{\Phi_L} T \ket{\Phi_R}  \sim \nu_0^{L_y}, 
\ee
with $\nu_0$ the dominant eigenvalue of the transfer matrix $E$. In terms of the left- and right-dominant eigenvectors of $E$ one finds that 
\be
\bra{\Phi_L} T \ket{\Phi_R}  \sim \left( \frac{\bra{\Sigma_U} E \ket{\Sigma_D}}{\braket{\Sigma_U}{\Sigma_D}}\right)^{L_y},  
\label{eq7}
\ee
with $\ket{\Sigma_U}$ and $\ket{\Sigma_D}$ the dominant left- and right- eigenvectors of $E$ respectively, which again we assume to be not necessarily normalized. This is shown in the tensor network diagram of Fig.~\ref{fig2}.

Now, let us focus on the denominator of Eq.~(\ref{fi}). Following a similar procedure as for the numerator, we realize that it is the product of two MPS, as shown in Fig.~\ref{fig1}. We have then that 
\be
\braket{\Phi_L}{\Phi_R} = {\rm tr} \left( K^{L_y} \right),  
\ee
with $K$ the 0D MPS transfer matrix shown in Fig.~\ref{fig3}. In the limit $L_y \gg 1$ we have then 
\be
\braket{\Phi_L}{\Phi_R} \sim  \theta_0^{L_y}, 
\ee
with $\theta_0$ the dominant eigenvalue of transfer matrix $K$. In terms of the dominant left- and right-eigenvectors of $K$, this can be written as 
\be
\braket{\Phi_L}{\Phi_R}  \sim \left( \frac{\bra{\Omega_U} K \ket{\Omega_D}}{\braket{\Omega_U}{\Omega_D}}\right)^{L_y}, 
\label{eq10} 
\ee
with $\ket{\Omega_U}$ and $\ket{\Omega_D}$ the dominant left and right eigenvectors of $K$ respectively, which again we assume to be not necessarily normalized. This is shown in the tensor network diagram of Fig.~\ref{fig3}.

\begin{figure}
	\centering
	\includegraphics[width=0.8\linewidth]{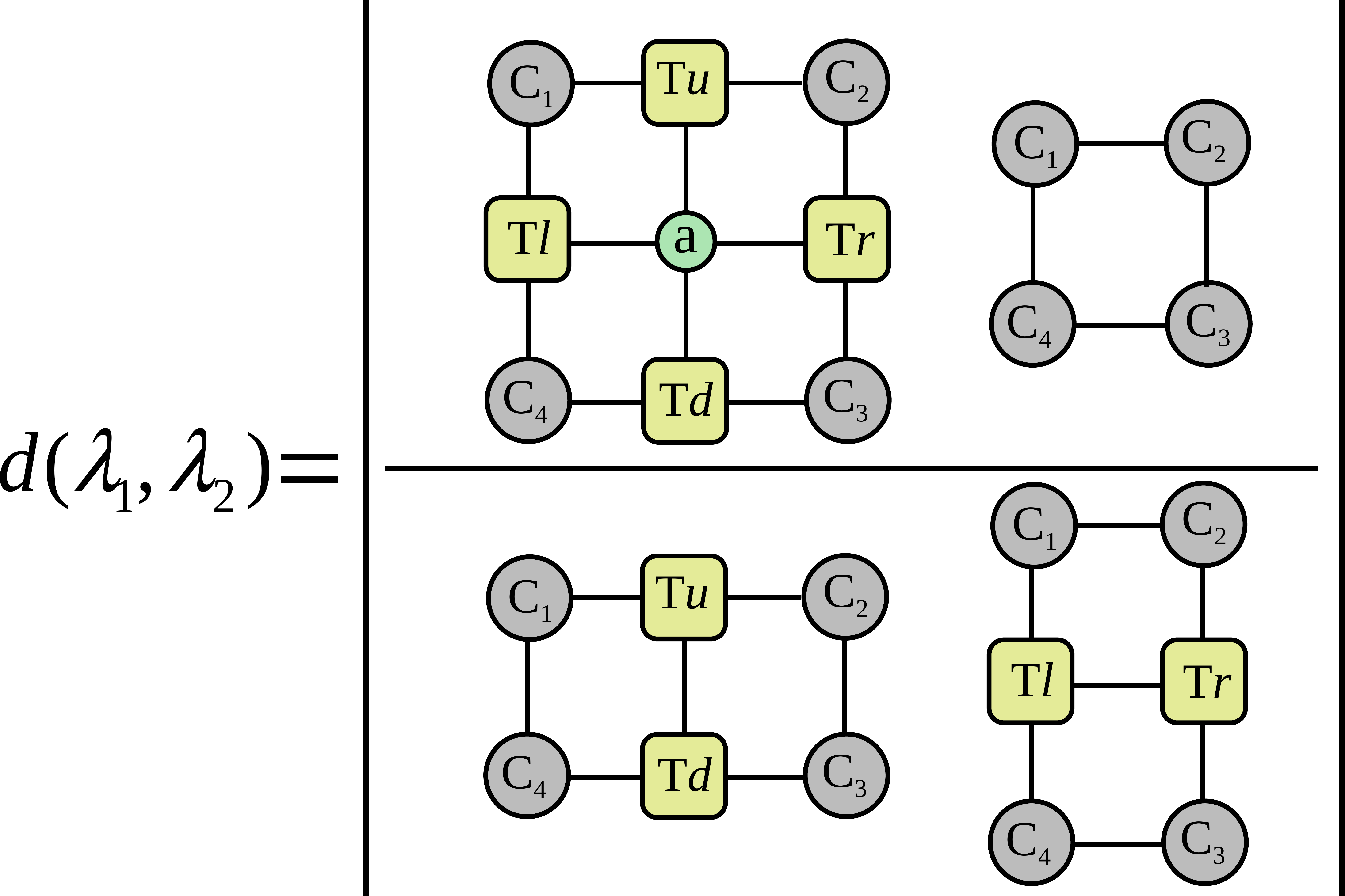}
	\caption{(Color online) Contraction providing $d(\lambda_1,\lambda_2)$, expressed in terms of the four CTMs $C_1, C_2, C_3$ and $C_4$, and the half-row and half-column transfer matrices $T_u, T_r, T_d$ and $T_l$.}
	\label{fig4}
\end{figure}

Putting everything together, we get the equation 
\be
F(\lambda_1, \lambda_2) \sim \left| \frac{\bra{\Sigma_U} E \ket{\Sigma_D} \braket{\Omega_U}{\Omega_D}}{{\braket{\Sigma_U}{\Sigma_D}\bra{\Omega_U} K \ket{\Omega_D}}} \right|^{L_x L_y}, 
\ee
which implies that 
\be
d(\lambda_1, \lambda_2) =\left|Ê \frac{\bra{\Sigma_U} E \ket{\Sigma_D} \braket{\Omega_U}{\Omega_D}}{{\braket{\Sigma_U}{\Sigma_D}\bra{\Omega_U} K \ket{\Omega_D}}} \right|.   
\label{fidelity}
\ee

The equation above is our main expression for the fidelity scaling variable $d(\lambda_1, \lambda_2)$, from which it is easy to extract the fidelity per site. The reason why Eq.~(\ref{fidelity}) is important, is that it admits an immediate interpretation in terms of CTMs, which we show in Fig.~\ref{fig4}. Notice that from the TN point of view, this is a very neat and clean expression, where we used the fact that all the dominant eigenvectors in Eq.~(\ref{fidelity}) can in fact be written, asymptotically and for an infinite lattice, in terms of the CTMs $C_1, C_2, C_3$ and $C_4$ as well as the half-row and half-column transfer matrices $T_u, T_r, T_d$ and $T_l$. Therefore, if one has a CTM algorithm at hand, one can also use it readily to compute Eq.~(\ref{fidelity}) as in Fig.~\ref{fig4} in order to get the fidelity per lattice site. 

\section{Boundary density operator from corner transfer matrices}
\label{Sec:BoundaryRho}

For the tensors obtained from the iPEPS algorithm, it is indeed possible to wrap them around an $N_h \times N_v$ cylinder and compute the entanglement entropy of half a cylinder in the limit $N_h \rightarrow \infty$. This is done using similar techniques to those in the calculation of the entanglement spectrum of PEPS, see Refs.~\cite{Li2008,Cirac2011,Yang2014,Poilblanc2016,Mambrini2016}. Here we wish to revise the essential ingredients of this calculation, and to show that it is indeed possible to do it using the tensors obtained from the CTM technique when contracting an infinite 2D lattice. 

\begin{figure*}
\centerline{\includegraphics[width=15cm]{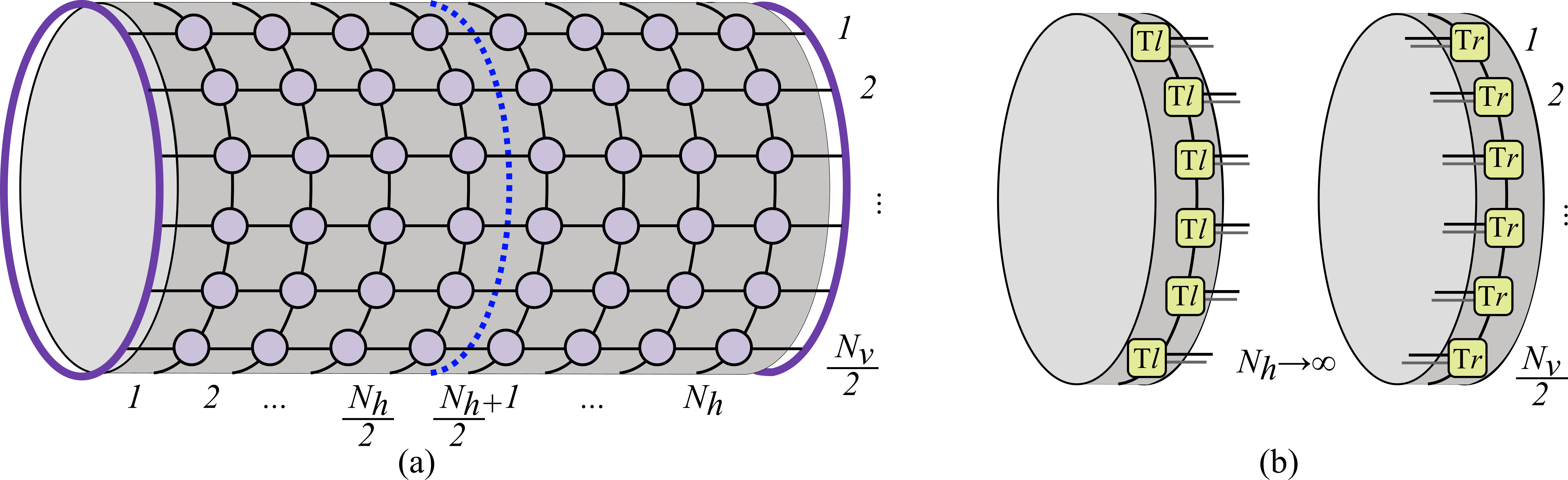}}
\caption{(Color online) (a) $N_h \times N_v$ cylinder, split into two halves by a blue dotted line. (b) The dominant left- and right- eigenvectors $\sigma_L$ (left) and $\sigma_R$ (right) of the PEPS transfer matrix on the cylinder.}
\label{Fig:cylinder}
\end{figure*}

Consider a 2D PEPS wrapped around a cylinder of circumference $N_v$, as in Fig.~\ref{Fig:cylinder}-(a), which we take to be infinitely-long. As shown in the figure, we split the cylinder in two parts (call them $L$ for ``left" and $R$ for ``right"). As explained in Ref.~\cite{Cirac2011}, the reduced density matrix of half an infinite cylinder, e.g., for $L$, is given by
\be
\rho = U \sqrt{\sigma_L^T}\sigma_R \sqrt{\sigma_L^T} U^\dagger,
\label{Eq:rho}
\ee
with $\sigma_{L/R}$ the reduced density operators in $L/R$ for the virtual 
spaces across the bipartition, and $U$ an isometry obtained from the contraction 
of the PEPS tensors. Mathematically, $\sigma_{L/R}$ corresponds to the  dominant 
left/right eigenvectors of the PEPS transfer matrix formed by the reduced 
tensors around the circumference of the cylinder. Using the above equation it is 
easy to see that $\rho$ has 
the same eigenvalues as $\sqrt{\sigma_L^T}\sigma_R \sqrt{\sigma_L^T}$, because 
the two operators are related by an isometry, thus leaving the eigenvalue 
spectrum invariant. Additionally, $\sqrt{\sigma_L^T}\sigma_R \sqrt{\sigma_L^T}$ 
turns out to have the same spectrum as $\sigma_L^T \sigma_R$ 
\cite{Poilblanc2016}. In order to compute the entanglement entropy of $\rho$, we 
can then focus on the eigenvalues of $\sigma_L^T \sigma_R$ exclusively. 

The dominant eigenvectors $\sigma_{L/R}$ are in fact easy to compute using CTM methods. As shown in Fig.~\ref{Fig:cylinder}-(b), these can be written entirely in terms of the half-row transfer matrix tensors $T_l$ and $T_r$ that are computed when approximating effective environments with CTMs. These tensors are then wrapped around a circle of length $N_v$, and constitutes our approximated $\sigma_{L/R}$. This approach is very efficient and provides accurate results. The diagonalization of  $\sigma_L^T \sigma_R$ then proceeds as usual, namely, using Krylov-subspace methods (e.g., Lanczos) which rely on matrix-vector multiplications. In our case such multiplications can be done very efficiently by exploiting the TN structure of  $\sigma_L^T \sigma_R$. From the approximated eigenvalues, the approximation to the EE just follows. 

\section{Ruby Model}

\label{Sec:model}

The ruby model, also known as two-body color-code model, was first introduced in 
Ref.~\cite{Bombin2009} as the first instance of a local Hamiltonian with 
two-body interactions, which reproduces the topological color-code model in the 
low energy sector of its gapped phase. The model supports string-net integrals 
of motion and respects the local and global $\Zd \times \Zd$ gauge 
symmetry in all of its limiting cases. The Hamiltonian of the ruby model (see 
Fig.\ref{Fig:ruby}-(a)) is defined as
\be
H_R=-\sum_{\alpha=x,y,z} J_{\alpha} \sum_{\alpha-\rm{links}} \sigma_i^{\alpha} \sigma_j^{\alpha} \quad,
\label{eq:H-KR}
\ee
where the first sum runs over $\alpha$-links ($\alpha=x,y,z$) labeled by red ($\mathrm{r}$), green ($\mathrm{g}$) and blue ($\mathrm{b}$) colors, respectively, and the second sum runs over 
the two-body interactions acting on sites $i$ and $j$ of the $\alpha-\rm{links}$, with $\sigma^{\alpha}$ the Pauli matrices. Without loss of generality, here we set $J_{\alpha}\!>\!0$.

Due to the four-valence structure of the ruby lattice, the ruby model is not exactly solvable and therefore an exact characterization of the underlying phases of the model is unavailable. Recently, an exact diagonalization (ED) study of the model on the 18 and 24 site ruby clusters detected three separate phases for the model in the $J_{x}+J_{y}+J_{z}=2$ plane, one of which is already known to be a robust \cite{Jahromi2013a, Jahromi2013, Capponi2014} gapped and topologically ordered \cite{Bombin2009,Kargarian2010} phase and the two others were conjectured to be new gapless spin-liquid phases. The low-energy effective theory of the gapless phases are further given by a local Hamiltonian on the triangular lattice \cite{Jahromi2016} with three-spin interactions. Since the effective Hamiltonians of the gapless phases are not exactly solvable, the characterization of the phases was performed with ED on finite clusters, which needs to be further explored.

In the following, we use the iPEPS method revisited in the previous sections for the triangle-honeycomb lattices in order to study the ruby model on an infinite 2D lattice, and extract its zero-temperature phase diagram directly in the thermodynamic limit. 

\section{Numerical results}
\label{Sec:Phase-Diag}

 \begin{figure}[t]
\centerline{\includegraphics[width=\columnwidth,trim={0.4cm 0 1.5cm 0},clip]{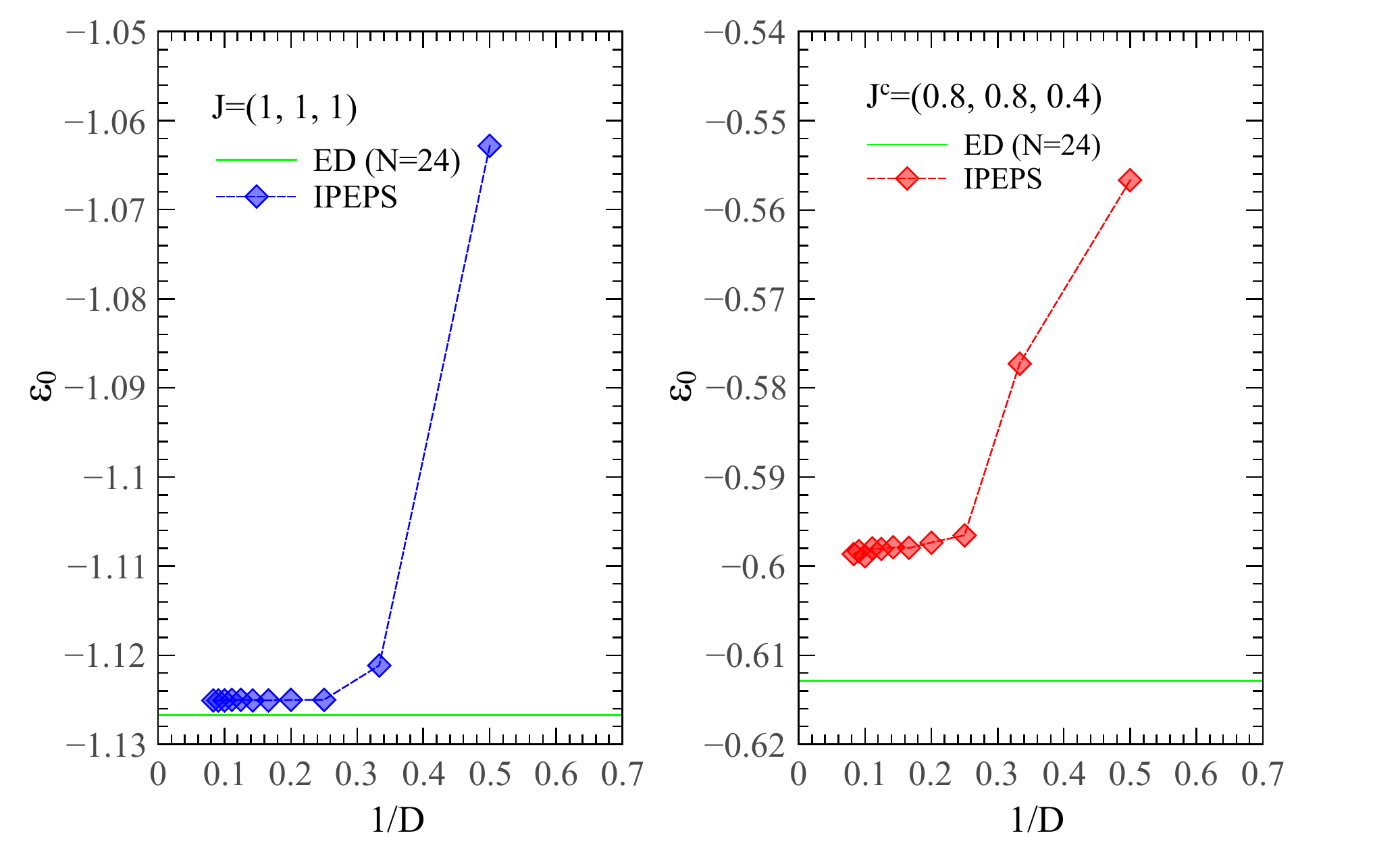}}
\caption{(Color online) Scaling of the iPEPS ground-state energy with respect to inverse bond dimension $D$. Left: Scaling of the energy for $(1,1,1)$ couplings compared to the ED results for $N=24$ sites.  Right: Scaling of the energy at the multicritical point, $\mathbf{J^c}=(0.8,0.8,0.4)$,  compared to the ED results for $N=24$ sites.    }
\label{Fig:e0scaling}
\end{figure}

In this section, we elaborate on the phase diagram of the ruby model and its 
possible quantum phase transitions by analyzing the ground-state properties of 
the system in different coupling regimes $(J_x,J_y,J_z)$. 

Before we start with the phase transitions, let us first benchmark the iPEPS energies  with the best available ED results (for 24 sites) \cite{Jahromi2016} at $(1,1,1)$ couplings. Fig.~\ref{Fig:e0scaling}-left shows the scaling of the ground-state energy per site, $\varepsilon_0$, for different bond dimensions $D$ compared to the ED result. As we can see, there is a very good agreement between the iPEPS energies for $D\geqslant 4$ and the ED energy $\varepsilon_0^{ED}=-1.12672$. In fact our best iPEPS energy, $\varepsilon_0=-1.125069$ for $D=12$ ($\chi=100$) is pretty close to that of the ED on finite ruby cluster with $N=24$ sites. We restricted the analysis of the phase diagram of the ruby model up to $D_{\rm Max}=8$. We further observed that going to higher bond dimensions would not change our findings, particularly away from the phase transitions. 

We also calculate the ground-state energy at the multicritical point, 
$\mathbf{J^c}=(0.8,0.8,0.4)$, for different bond dimension $D$. 
Fig.~\ref{Fig:e0scaling}-right depicts the scaling of energy per site versus 
inverse bond dimension compared to the ED ($N=24$). In contrast to the isotropic 
case ($1,1,1$), the best energy obtained with iPEPS ($D=12$) is still higher than the one 
obtained by ED on a 24-site cluster. This may be due to the large amount of correlations present at this point, which affect, at the same time but in 
different ways, both the ED and iPEPS calculations, leading to a difference 
between both energies of around $2 \% -  3\%$. Our best variational iPEPS energy 
in this case is $\varepsilon_0=-0.59861$ for $D=12$ ($\chi=100$), which is 
slightly higher than the one obtained by ED for 24 sites, 
$\varepsilon_0^{ED}=-0.61285$. 

 \begin{figure}[t]
\centerline{\includegraphics[width=\columnwidth]{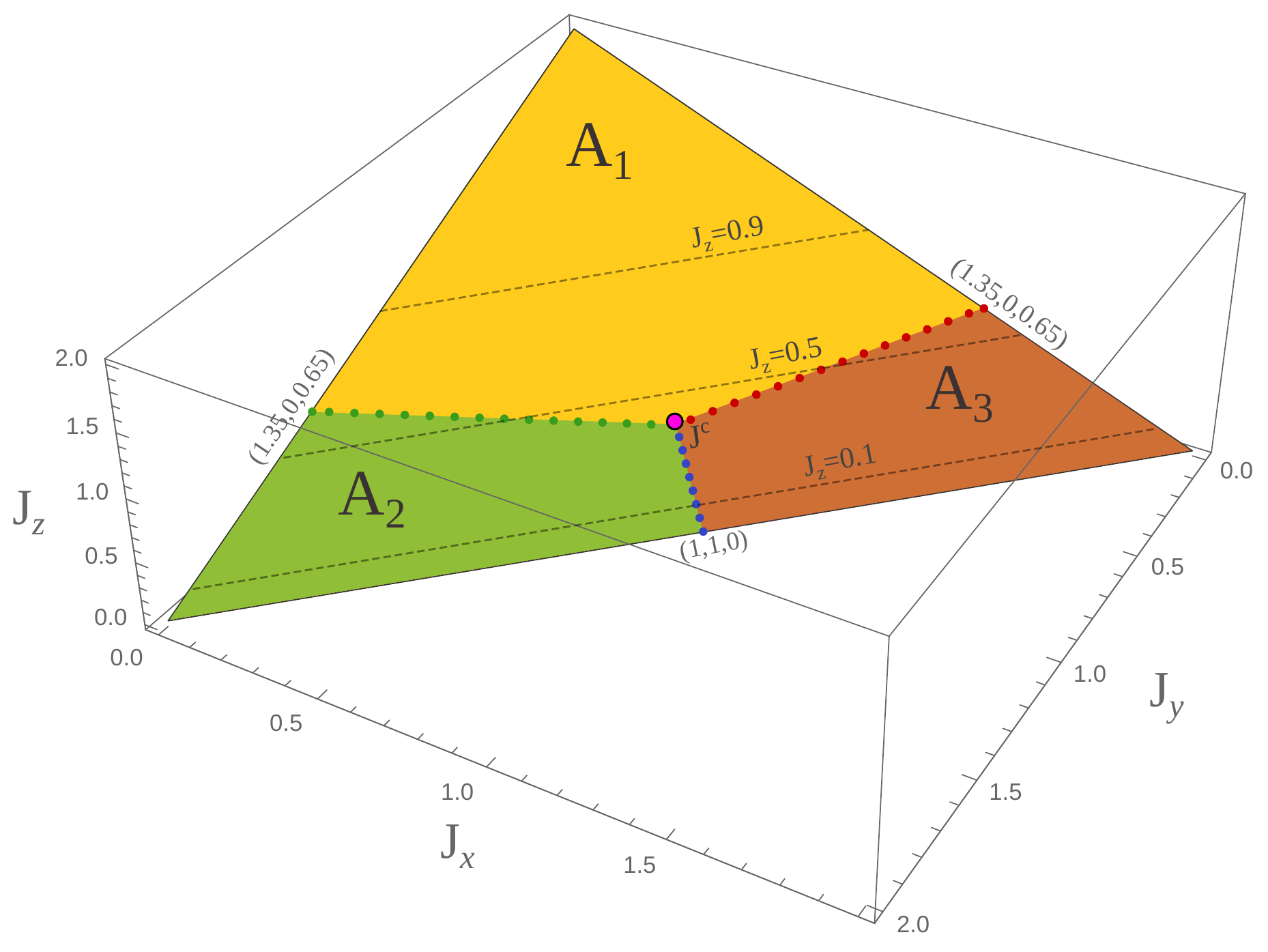}}
\caption{(Color online) Phase diagram of the ruby model on an infinite 2D lattice obtained with the iPEPS technique.  The phase diagram is composed of three distinct phases, $A_1$, $A_2$ and $A_3$, which are separated from each other by second-order phase transition lines meeting at a multicritical point $\mathbf{J^c}=(0.8,0.8,0.4)$. The phase diagram confirms previous findings with ED on finite-size clusters \cite{Jahromi2016}.}
\label{Fig:phasediag}
\end{figure}

In order to study the phase diagram of the ruby model and capture the possible phase transitions, we restricted the couplings to the $J_x+J_y+J_z=2$ plane and approximated the ground state of the model on the parameter lines $J_x+J_y=2-J_z$ with fixed $J_z$ throughout the whole plane for $0 \leqslant J_z\leqslant 2$. These parameter lines are labeled with $J$ in the forthcoming figures. We captured possible phase transitions by evaluating different observables such as the nearest-neighbor correlators $\expectval{\sigma_i^\alpha \sigma_j^\alpha}$ ($\alpha=x,y,z$), the ground-state fidelity $F(\lambda_1, \lambda_2) = |\braket{\Psi(\lambda_2)}{\Psi(\lambda_1)}|$, and the EE on an infinite cylinder. 

\begin{figure*}
 \centering
 \begin{tabular}{cc}
 \includegraphics[width=0.5\textwidth,trim={0 0 0 0},clip]{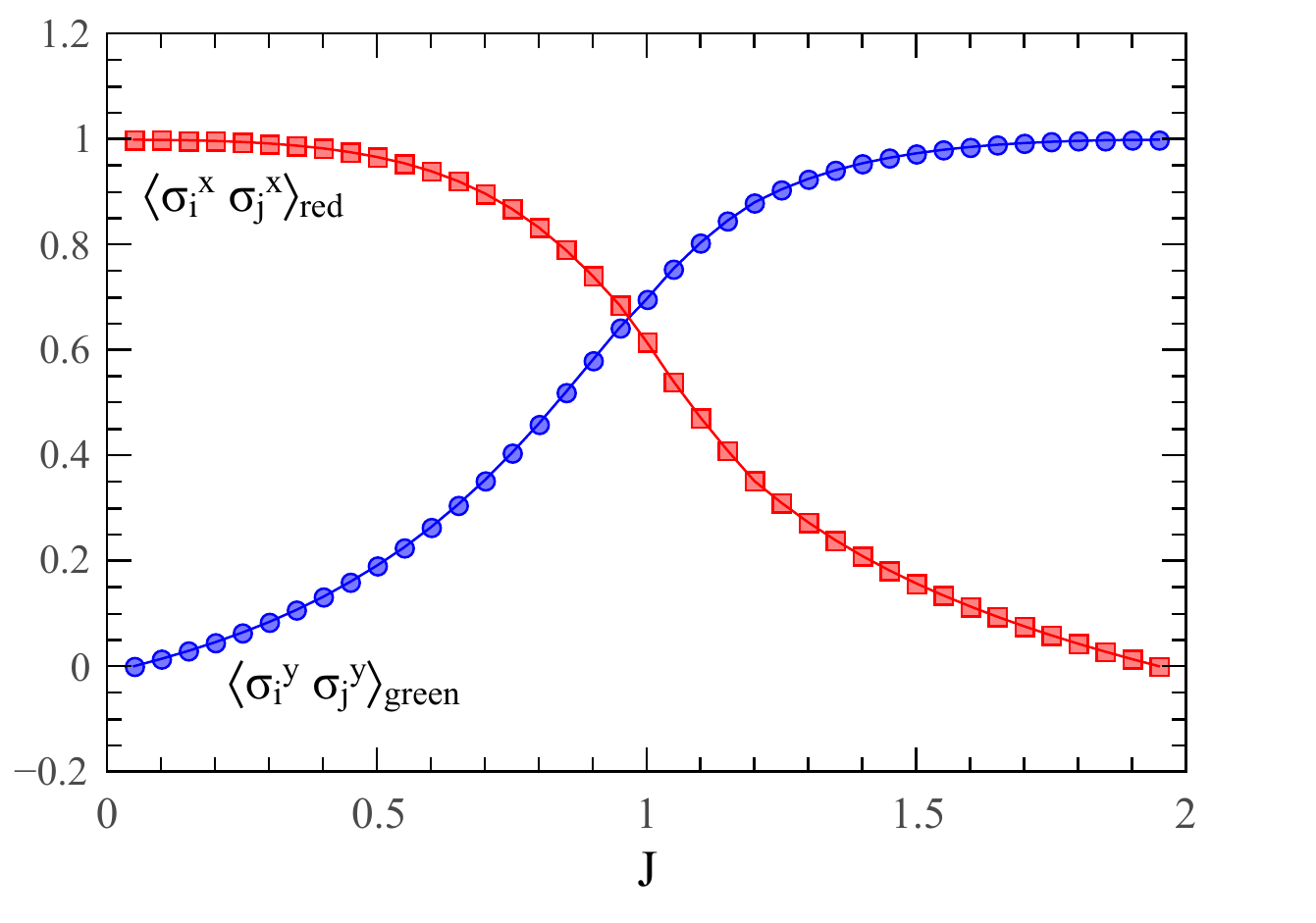} & \includegraphics[width=0.5\textwidth,trim={0 0 0 0},clip]{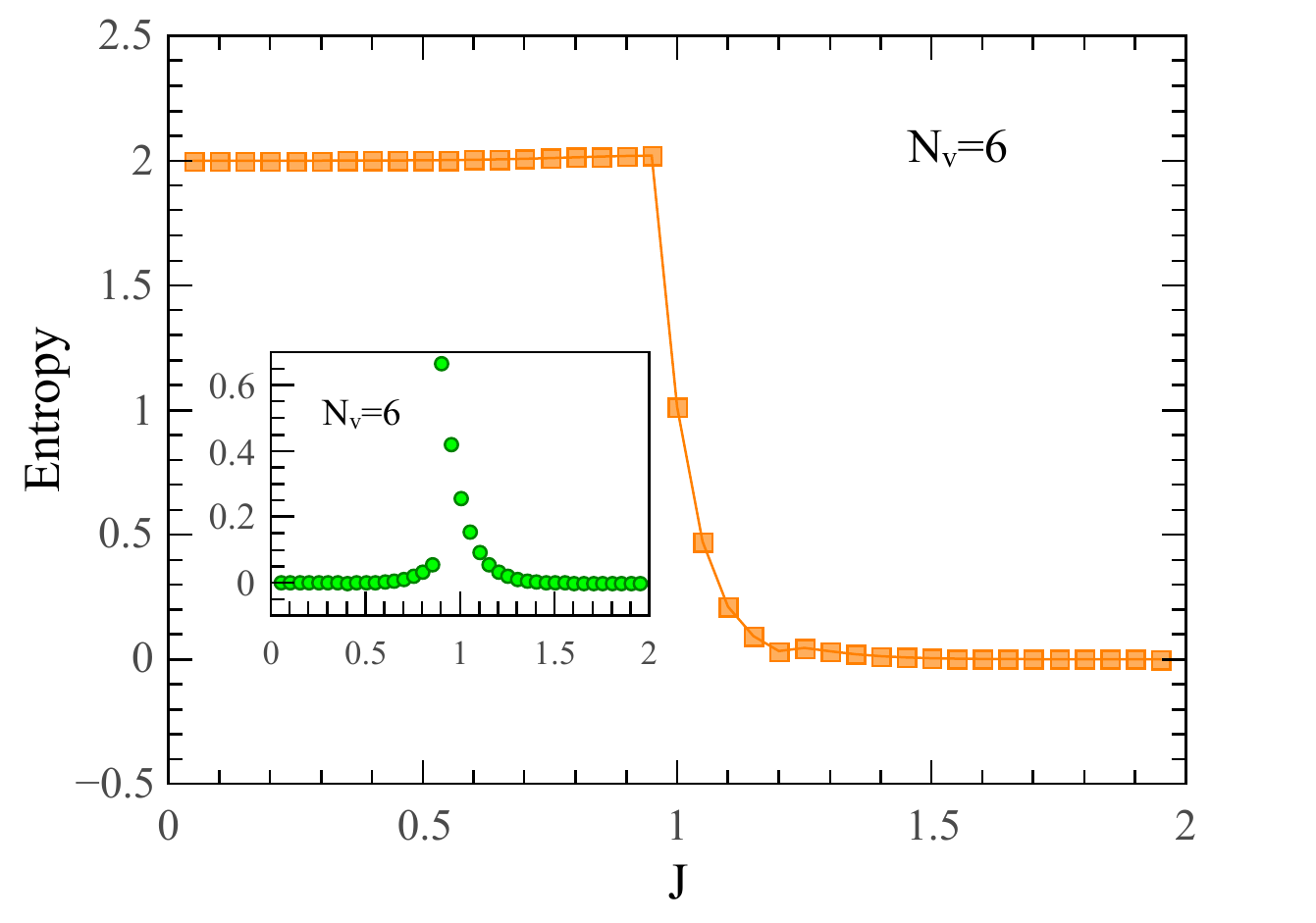} \\
 (a) & (b) 
 \end{tabular}
 \caption{(Color online) (a) Two-point correlators $\expectval{\sigma_i^x \sigma_j^x}$ and $\expectval{\sigma_i^y \sigma_j^y}$ on red and green links, respectively and (b) $S^{VN}$ on a cylinder with $N_h=\infty, N_v=6$ obtained with simple update, along the 1D line $J_x+J_y=2-J_z$ for fixed $J_z=0.1$ and varying $J_x, J_y$. We show this line by $J$. The results are for $D=8$ and $\chi = 80$. We observed that (in contrast to conventional systems such as Ising models) EE in the ruby model is highly sensitive to the iPEPS minimization algorithm and initial conditions of the tensors. The inset in (b) depicts the EE obtained with full update algorithm for $D=3$ and different initial condition for iPEPS tensors which is more consistent with the nature of $A2$, $A3$ phases which are identical up to the interchange of $J_x$, $J_y$ couplings and therefore should have the same EE.}
 \label{Fig:jz01} 
 \end{figure*}
 
\begin{figure}[t]
\centerline{\includegraphics[width=\columnwidth,trim={2.5cm 7cm 3cm 7cm},clip]{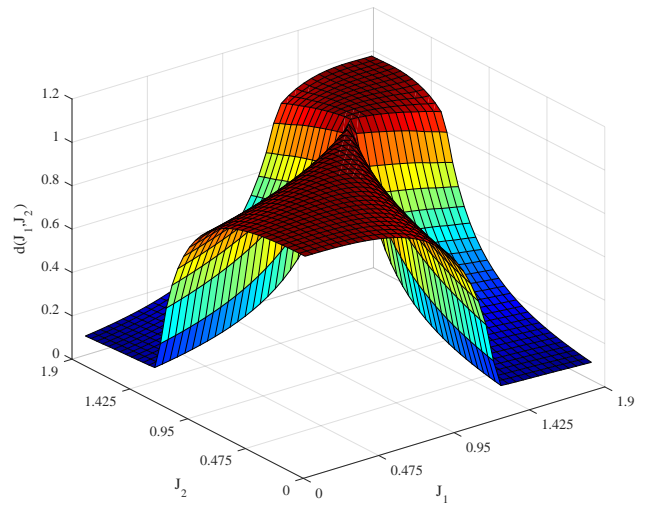}}
\caption{(Color online) 3D surface plot of the ground-state fidelity per lattice site, $d(J_1,J_2)$, along the 1D line $J_x+J_y=2-J_z$ for fixed $J_z=0.1$ and varying $J_x, J_y$. We show this line by $J$. $J_1$ and $J_2$ denote different points on the parameter line $J$.}
\label{Fig:fidelity01}
\end{figure}

The computed phase diagram of the ruby model on the $J_x+J_y+J_z=2$ plane is shown in Fig.~\ref{Fig:phasediag}. This is composed of three distinct phases $A_1$, $A_2$ and $A_3$ which are separated from each other by second order phase transition lines that meet at a multicritical point $\mathbf{J^c}=(0.8,0.8,0.4)$. In order to capture the phase transitions, we scanned the plane along the fixed $J_z$ lines starting from $J_z=0$ and pinpointed the transition points by evaluating different observables. Fig.~\ref{Fig:jz01} shows the correlators $\expectval{\sigma_i^x \sigma_j^x}$ and $\expectval{\sigma_i^y \sigma_j^y}$ on red and green links, respectively as well as the Von Neumann EE ($S^{VN}$) of half an infinite cylinder for $J_z=0.1$. As we can see, the smooth change of correlation between zero and one signals a continuous phase transition at the $J_x=J_y$ point which proves the existence of two distinct phases i.e., $A_2$ and $A_3$. This is further confirmed by the surface plot of the ground state fidelity which, independently of the nature of the underlying phases, is a powerful probe for capturing the phase transition. Fig.~\ref{Fig:fidelity01} depicts the ground-state fidelity for the ruby model for $J_z=0.1$. The points on the fidelity surface plot were calculated as an overlap between the ground state wave function for two different couplings $J_1, J_2$ located on the parameter line $J$ (see definition for $J$ in previous paragraph). The continuous change of fidelity surface plot at $J_1\approx1$ is a signature of a second order phase transition \cite{Zhou2008} between $A_2$ and $A_3$ phases. Let us further note that the results obtained for small bond dimensions, $D$, shows discontinuities in both correlations and fidelity. However, by increasing the bond dimension the curves become continuous and the results are indeed in favor of a second order phase transition in the $D\longrightarrow\infty$ limit.

\begin{figure*}
 \centering
 \begin{tabular}{cc}
 \includegraphics[width=0.5\textwidth,trim={0 0 0 0},clip]{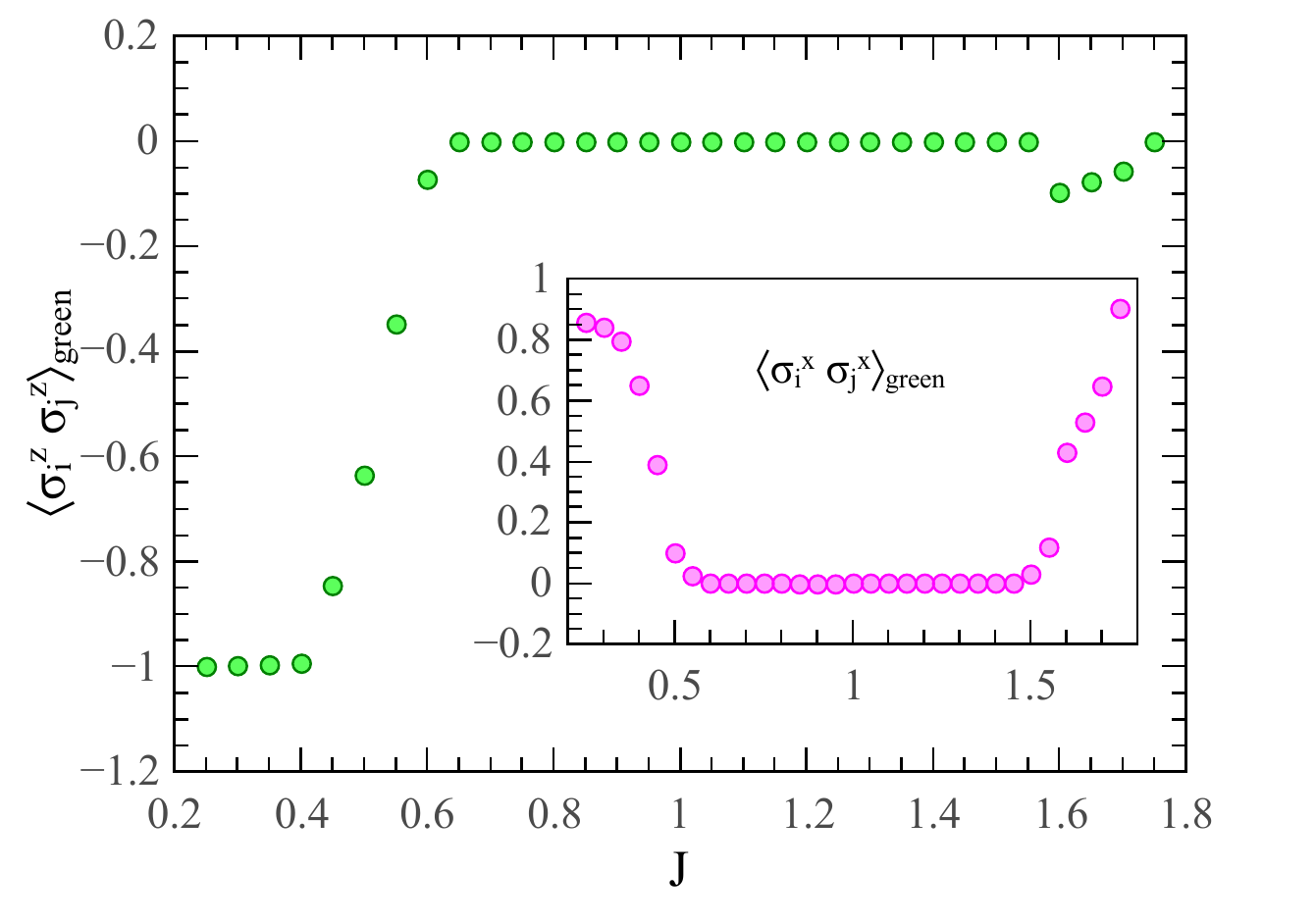} & \includegraphics[width=0.5\textwidth,trim={0 0 0 0},clip]{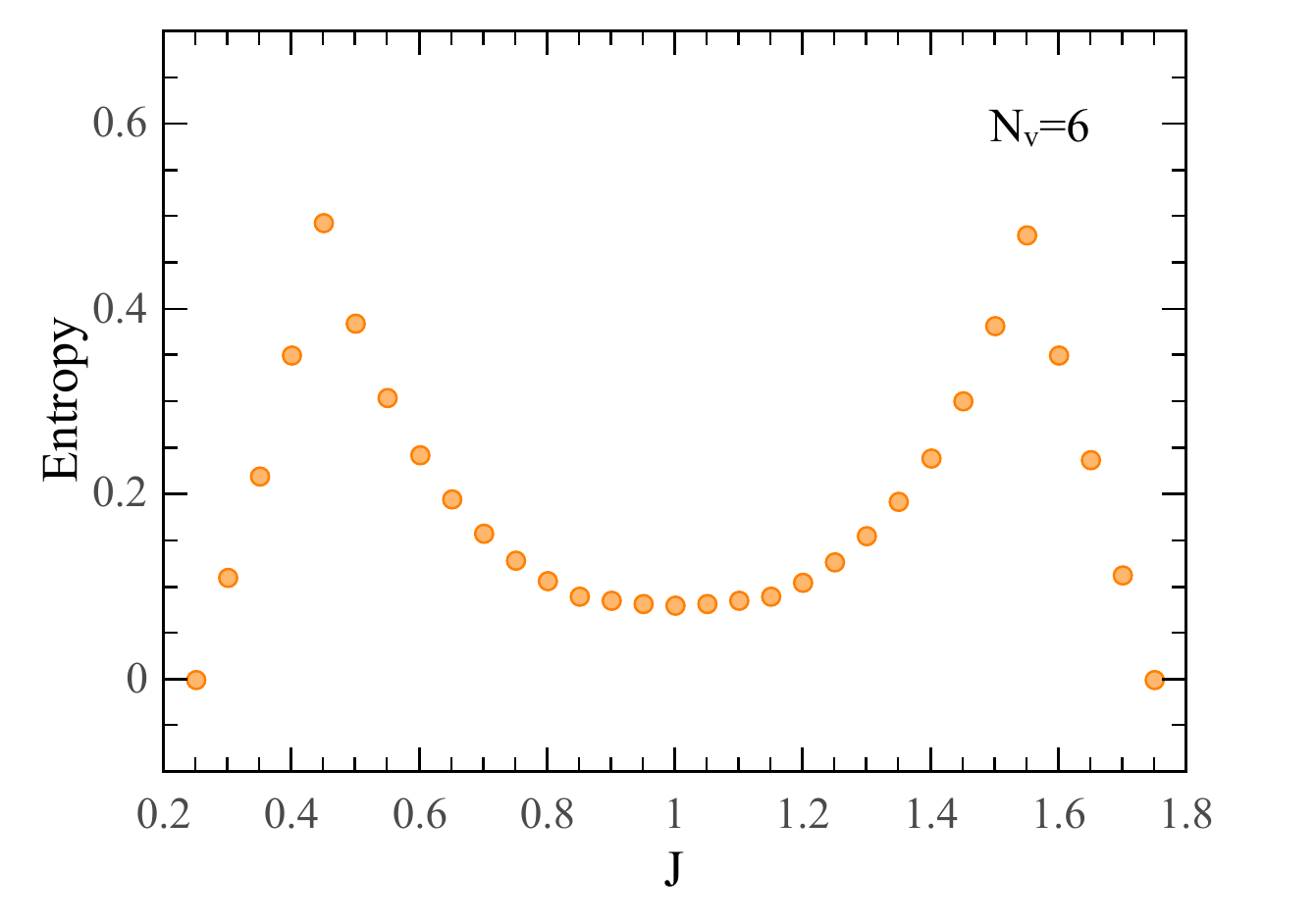} \\
 (a) & (b) \\
 \end{tabular}
 \caption{(Color online) (a) Two-point correlator $\expectval{\sigma_i^z \sigma_j^z}$ and $\expectval{\sigma_i^x \sigma_j^x}$ (inset) on green links. (b) $S^{VN}$ on a cylinder with ($N_h=\infty, N_v=6$) along the 1D line $J_x+J_y=2-J_z$ for fixed $J_z=0.5$ and varying $J_x, J_y$. We show this line by $J$. The results are for $D=8$ and $\chi = 80$.}
 \label{Fig:jz05}
 \end{figure*}

This behavior continues until $J_z=0.4$, above which we start to capture two phase transitions, thus signaling the existence of three distinct phases. Fig.~\ref{Fig:jz05} shows different local observables along the parameter line with $J_z=0.5$. As we can see from entropy and two-point correlators on different links, there are two discontinuities in the figures:  the left one confirms the transition from the $A_2$ phase into $A_1$, and the right one the transition from the $A_1$ phase into $A_3$. These two transition points are also captured for $0.4 \leqslant J_z\leqslant 0.65$.

\begin{figure*}
 \centering
 \begin{tabular}{cc}
  \includegraphics[width=0.5\textwidth,trim={0 0 0 0},clip]{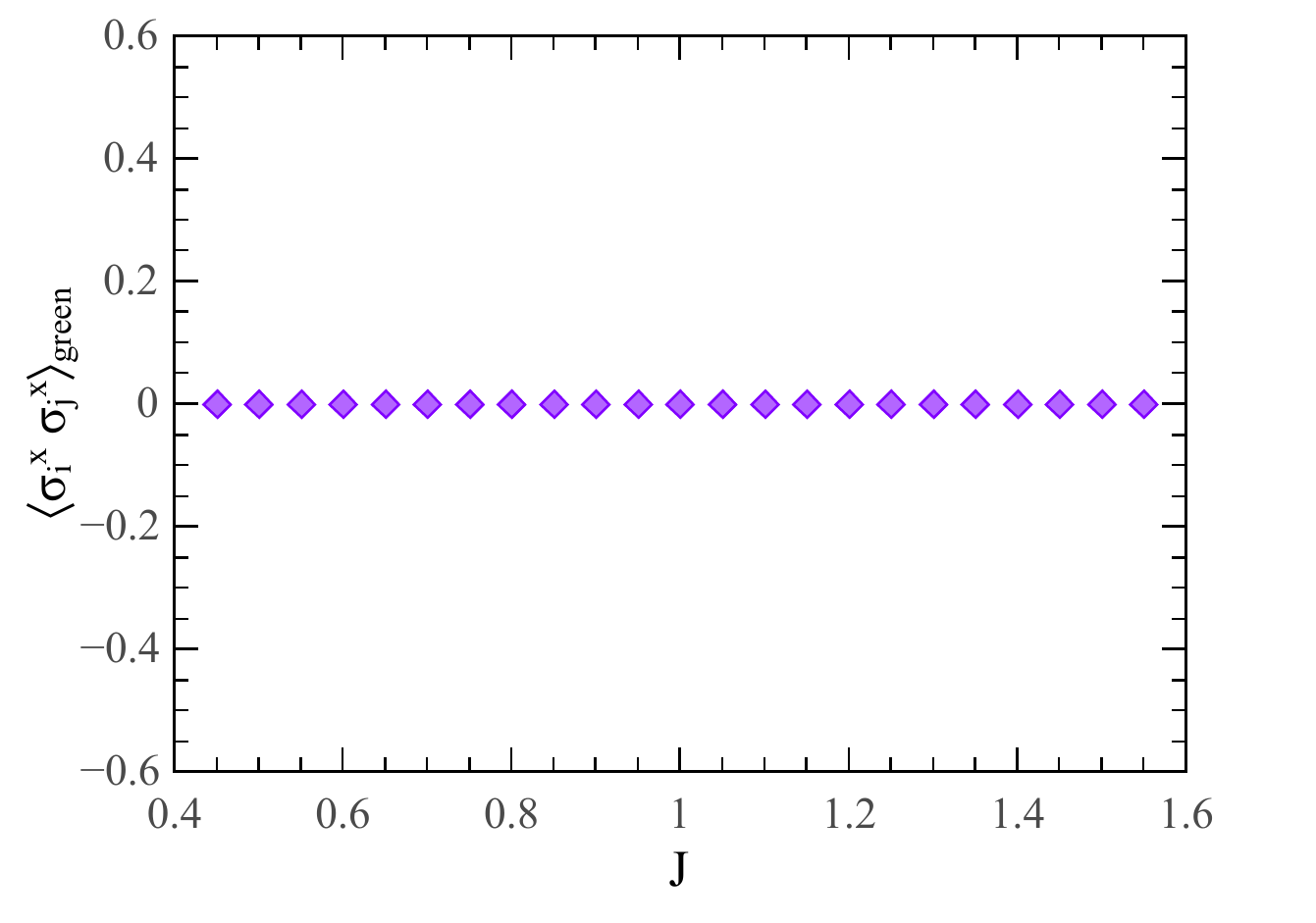} & \includegraphics[width=0.5\textwidth,trim={0 0 0 0},clip]{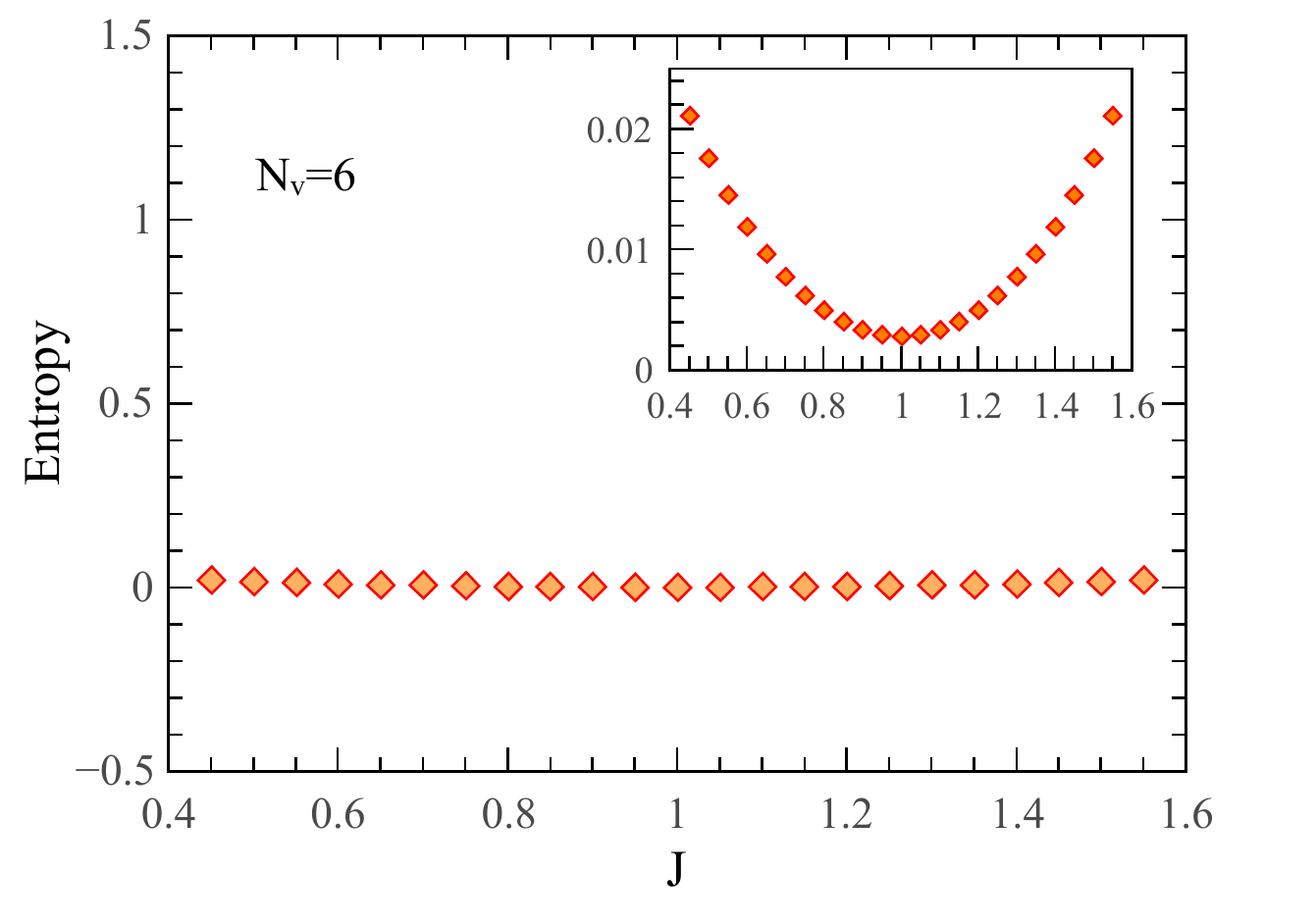}\\
 (a) & (b) 
 \end{tabular}
 \caption{(Color online) Two-point correlator $\expectval{\sigma_i^x \sigma_j^x}$ on green links and (b) $S^{VN}$ on a cylinder with $N_h=\infty, N_v=6$ along the 1D line $J_x+J_y=2-J_z$ for fixed $J_z=0.9$ and varying $J_x, J_y$. We show this line by $J$. The results are for $D=8$ and $\chi = 80$.}
  \label{Fig:jz09}
 \end{figure*}

By increasing $J_z\geqslant 0.65$, the discontinuities in the local observables disappear and we no longer detect any phase transition. This means that we are inside the same phase i.e. the $A_1$ phase, in the whole range of couplings with $J_z\geqslant 0.65$. Fig.~\ref{Fig:jz09} shows $\expectval{\sigma_i^x \sigma_j^x}$ on the green links and $S^{VN}$ for $J_z=0.9$. The plots certify that no phase transition is detected and the system remains in the $A_1$ phase. This is further confirmed by calculations of the ground-state fidelity (not shown), whose diagram turns into a flat surface indicating no change in the ground-state of the system for different $J_x$ and $J_y$ couplings.

\begin{figure}
\centerline{\includegraphics[width=\columnwidth,trim={0 0 0 0},clip]{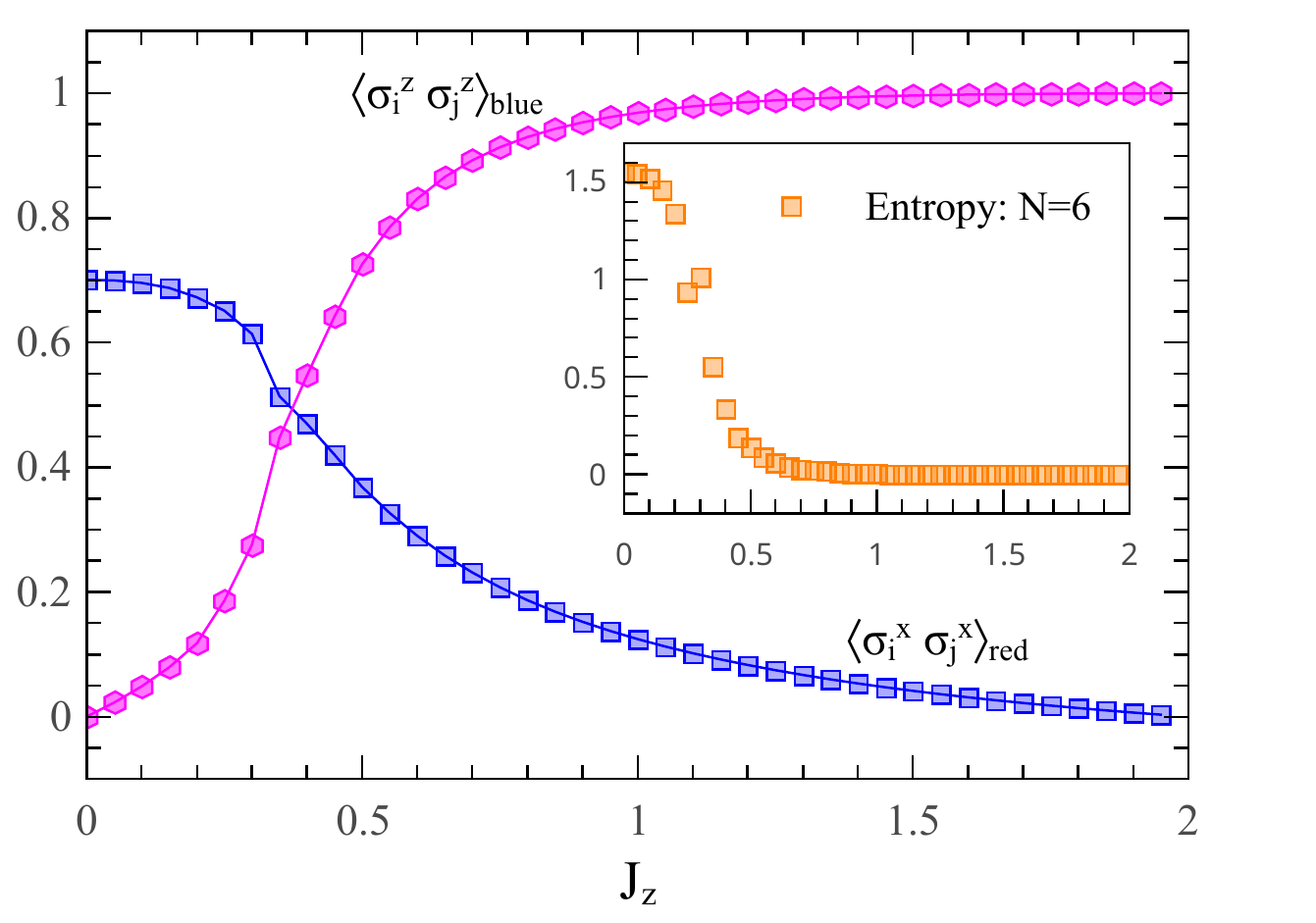}}
\caption{(Color online) Two-point correlator $\expectval{\sigma_i^x \sigma_j^x}$ and $\expectval{\sigma_i^z \sigma_j^z}$ on red and blue links, respectively as well as $S^{VN}$ for $N=6$ (inset) along the 1D line $J_z=2-J_x-J_y$ with $J_x=J_y$. The results are for $D=8$ and $\chi = 80$.}
\label{Fig:jzxx}
\end{figure}

\begin{figure} 
\centerline{\includegraphics[width=\columnwidth,trim={2.7cm 7cm 3cm 7.5cm},clip]{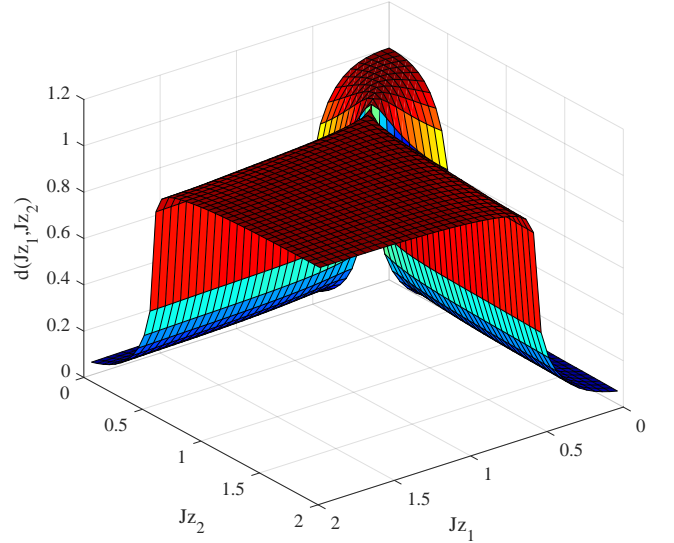}}
\caption{(Color online) Surface plot of the ground state fidelity per lattice site, $d(J_{z_1},J_{z_2})$, 
along the 1D line $J_z=2-J_x-J_y$ with $J_x=J_y$. $J_{z_1}$ and $J_{z_2}$ denote different points on the parameter line with varying $J_z$.}
\label{Fig:fidelityjz}
\end{figure}

In order to cross-check the existence of no other phase transitions for $J_z\geqslant 0.65$, which is a large region of the $J_x+J_y+J_z=2$ plane, and to further locate the multicritical point more accurately, we scanned the phase plane along the line $J_z=2-J_x-J_y$ with $J_x=J_y$ and calculated different quantities. Fig.~\ref{Fig:jzxx}) demonstrates the $\expectval{\sigma_i^x \sigma_j^x}$ and $\expectval{\sigma_i^z \sigma_j^z}$ on red and blue links, respectively as well as $S^{VN}$ for $N=6$ (inset) along this scan line. Our results once again confirm that, for a large region corresponding to the $A_1$ phase, we capture no phase transition until we reach the multicritical point at $(0.8,0.8,0.4)$. The 3D fidelity surface plot along this scan line, Fig.~\ref{Fig:fidelityjz}, also confirms that there is no phase transition for a large region inside the $A_1$ phase (see the large plateau for large $J_z$ couplings).

To conclude this section, let us mention that the resulting phase diagram from the iPEPS technique in the thermodynamic limit confirms previous findings with ED on finite-size clusters \cite{Jahromi2016}.

\section{Discussion and conclusion}
\label{Sec:Conclude}

In this paper, we have developed the machinery of the iPEPS algorithm with CTMs in order to apply it to the family of triangle-honeycomb structures, such as ruby and star lattices. We prescribed how the local Hilbert space of the triangles on the lattice can be replaced with block sites, and how one can implement the iPEPS method with the CTM algorithm accordingly. Furthermore, we showed how the CTM method can be used to calculate the ground-state fidelity per lattice site and the boundary density operator on infinite cylinders, which are powerful probes to be used for studying the ground-state properties of a given system and for capturing quantum phase transitions.

In order to examine the efficiency and accuracy of our iPEPS algorithm, we applied the method to the ruby model and investigated the phase diagram of the model in different coupling regimes in the thermodynamic limit. We found that the phase diagram of the ruby model on the $J_x+J_y+J_z=2$ plane is composed of three distinct phases, i.e., the $A_1$, $A_2$ and $A_3$ phases which are separated from each other by continuous phase transition points meeting at a multicritical point $\mathbf{J^c}=(0.8,0.8,0.4)$. The phase boundaries were captured by analyzing two-point correlators, ground-state fidelity and entanglement entropy of half an infinite cylinder. The $A_1$ phase is already known to be a gapped topological phase, whose low-energy physics is given by the effective topological color-code on the honeycomb lattice. The $A_2$ and $A_3$ phases are new phases which were first detected with ED on finite-size lattices with $18$ and $24$ sites \cite{Jahromi2016} and predicted to be gapless spin-liquids.


The new iPEPS phase diagram is in full agreement with our previous findings in Ref.~\cite{Jahromi2016}. However, we were not successful in capturing topological characteristics of the underlying phases of the model such as topological entropy \cite{Kitaev2006a,Levin2006}, $\gamma$, and modular matrices \cite{Zhang2012}, containing the anyonic statistics of quasiparticles, even for the $A_1$ phase which is already known to be $\Zd\times\Zd$ topologically ordered with $\gamma=2$ and abelian statistics. \cite{Kargarian2008,Jahromi2017}. The reason for this is that the detection of topological order from scratch (i.e. without some previous knowledge of the underlying gauge symmetry) is quite difficult in the context of current iPEPS techniques. As such, the current iPEPS algorithm does not a priori respect any gauge symmetry, and therefore has a hard time capturing any emergent gauge symmetry in the low-energy sector of a Hamiltonian. This, in turn, implies that information regarding the topological invariants may be lost in the optimization procedure, in spite of getting a very accurate description of the ground state energy as well as other local observables. We believe, however, that such a gauge-invariant optimization may indeed be possible, and therefore leave the door open for further developments in this respect.  

\section{Acknowledgements}
The authors acknowledge H. Yarloo and A. Kshetrimayum for helpful discussions on tensor network algorithms.
S.S.J. also acknowledges R.O. for hospitality during his stay at the Johannes Gutenberg University (JGU).
S.S.J. and A.L. acknowledge the support from the Iran Science Elites Federation (ISEF) and 
the Sharif University of Technology's (SUT) Office of Vice President for Research.
The iPEPS calculations were performed on the HPC cluster of SUT and Mogon cluster at JGU.

\appendix

\section{${\rm i}$PEPS implementation of the ruby Hamiltonian}
\label{appx:KR-IPEPS}
\begin{figure}
\centerline{\includegraphics[width=0.8\columnwidth]{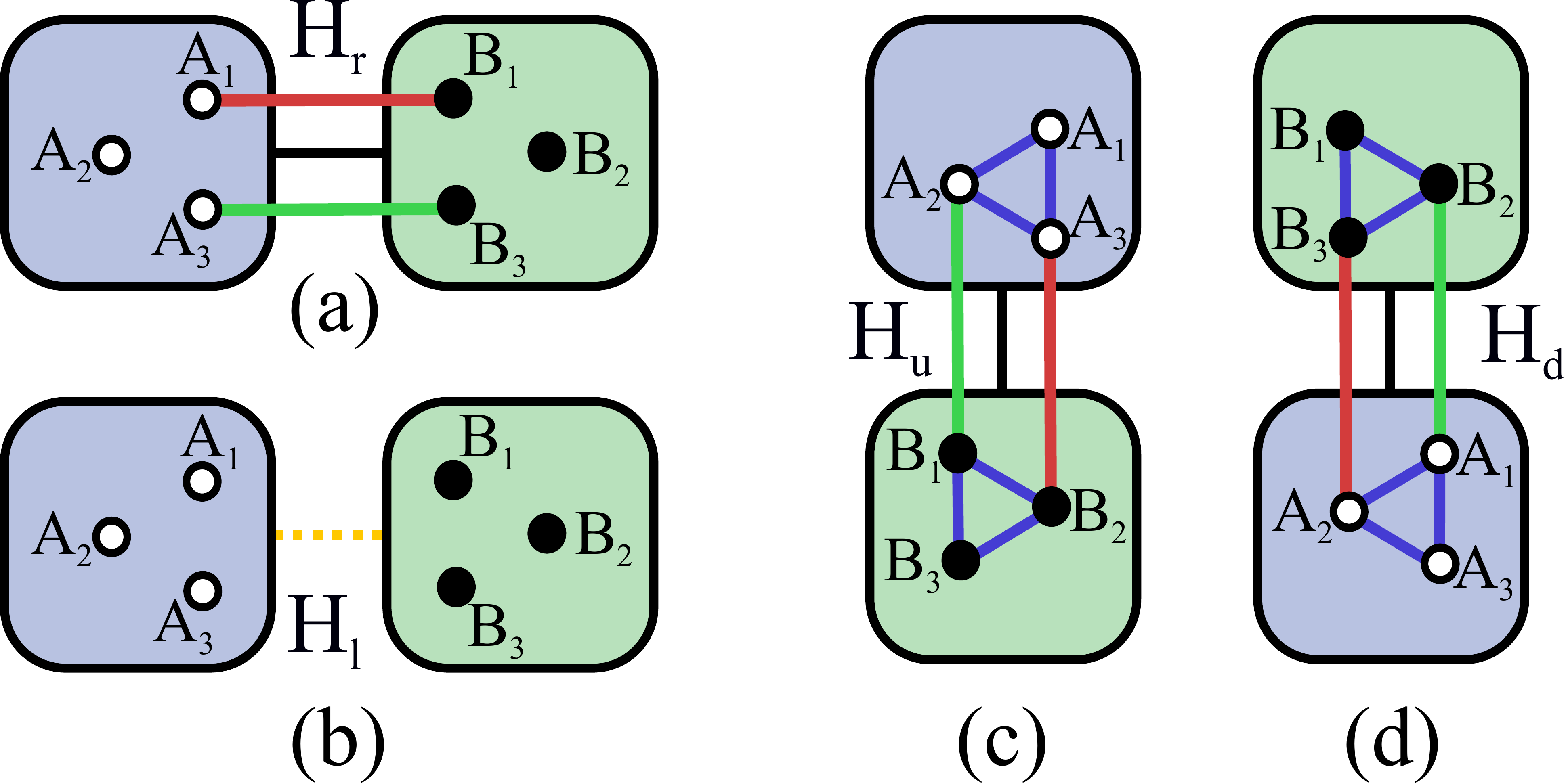}}
\caption{(Color online) Two-body local terms of the ruby model which act on nearest-neighbor block sites, and are further used in updating the $(l,r,u,d)$ bond indices of the $A$, $B$ iPEPS tensors. (a) $H_r$, which acts on two nearest-neighbor block sites along the $x$-direction. (b) $H_l$, which acts trivially i.e., as identity, on two nearest-neighbor block sites along the $x$-direction. (c) $H_u$, which is a summation of two local terms which act individually on each of the block sites as well as nearest-neighbor terms along the $y$-direction. (d) $H_d$, which is a summation of two local terms which act individually on each of the block sites as well as nearest-neighbor terms along the $y$-direction. See also Appendix.~\ref{appx:KR-IPEPS} for the explicit operator form of the $H_i$ terms.}
\label{Fig:hamiltonian}
\end{figure}

In this section, we explain how to map the nearest-neighbour interactions on the ruby lattice, Hamiltonian \eqref{eq:H-KR}, to the nearest-neighbour interactions on the square lattice (see also Ref.~\cite{Corboz2012a} for the kagome lattice). In the main text, we pointed out how one can reduce the ruby lattice to a brick-wall honeycomb lattice by grouping the spins on the vertices of each blue triangles in the ruby unit cell into two distinct block sites $A$, $B$ (see Fig.~\ref{Fig:ruby}-(c)). Labeling the spins in the block site $A$ ($B$) by $A_1$, $A_2$, $A_3$ ($B_1$, $B_2$, $B_3$) according to Fig.~\ref{Fig:hamiltonian}, the local Hilbert space of the a ruby unit cell (or equivalently two neighbuoring block sites) is given by
\bea
\Hi_{\rm cell}&=&\Hi_A \otimes \Hi_B, \\
\Hi_A &=& \Hi_{A_1}\otimes \Hi_{A_2} \otimes \Hi_{A_3}, \\
\Hi_B &=& \Hi_{B_1}\otimes \Hi_{B_2} \otimes \Hi_{B_3},
\eea
where the local physical dimension of $\Hi_A$ and $\Hi_B$ is $2^3=8$. The ruby model can therefore be represented in terms of local and nearest-neighbour interactions among the block sites as follows
\be
H_R=\sum_{i} H_i +\sum_{\langle i,j\rangle} H_{i,j}^x +\sum_{\langle i,j\rangle} H_{i,j}^y,
\ee
where
\bea
H_i&=& h_{iA}+h_{iB}, \\
H_{i,j}^x&=& h_{iA,jB}^x+h_{iB,jA}^x, \\
H_{i,j}^y&=& h_{iA,jB}^y+h_{iB,jA}^y,
\eea
with local terms 
\bea
h_{iA}&=& \sigma_{A_1}^z \sigma_{A_2}^z+\sigma_{A_1}^z \sigma_{A_3}^z+\sigma_{A_2}^z \sigma_{A_3}^z,\\
h_{iB}&=& \sigma_{B_1}^z \sigma_{B_2}^z+\sigma_{B_1}^z \sigma_{B_3}^z+\sigma_{B_2}^z \sigma_{B_3}^z, 
\eea
and nearest-neighbor interactions in the $x$-direction
\bea
H_{i,j}^x&=& h_{iA,jB}^x+h_{iB,jA}^x, \\
h_{iA,jB}^x&=& \sigma_{A_1}^x \sigma_{B_1}^x+\sigma_{A_3}^y \sigma_{B_3}^y,\\
h_{iB,jA}^x&=& \mathbb{I},
\eea
where $\mathbb{I}$ is the identity operator. Besides, the nearest-neighbour interactions in the $y$-direction are given by
\bea
H_{i,j}^y&=& h_{iA,jB}^y+h_{iB,jA}^y, \\
h_{iA,jB}^y&=& \sigma_{A_3}^x \sigma_{B_2}^x+\sigma_{A_2}^y \sigma_{B_1}^y,\\
h_{iB,jA}^y&=& \sigma_{B_3}^x \sigma_{A_2}^x+\sigma_{B_2}^y \sigma_{A_1}^y.
\eea

Eventually, the local two-body terms $H_i$, $i\in$ $(r,l,u,d)$ which are used in the imaginary time evolution process in  iPEPS are given in terms of interactions between block sites as  
\bea
H_r &=& h_{iA,jB}^x, \\
H_l &=& h_{iB,jA}^x, \\
H_u &=& \frac{1}{2} (h_{iA}+h_{iB})+h_{iA,jB}^y,\\
H_d &=& \frac{1}{2} (h_{iB}+h_{iA})+h_{iB,jA}^y.
\eea
The explicit definition of $H_i$ terms have also depicted in Fig.~\ref{Fig:hamiltonian}.

\section{Block structure of the star lattice}
\label{appx:star-block}

\begin{figure}[t]
\centerline{\includegraphics[width=\columnwidth]{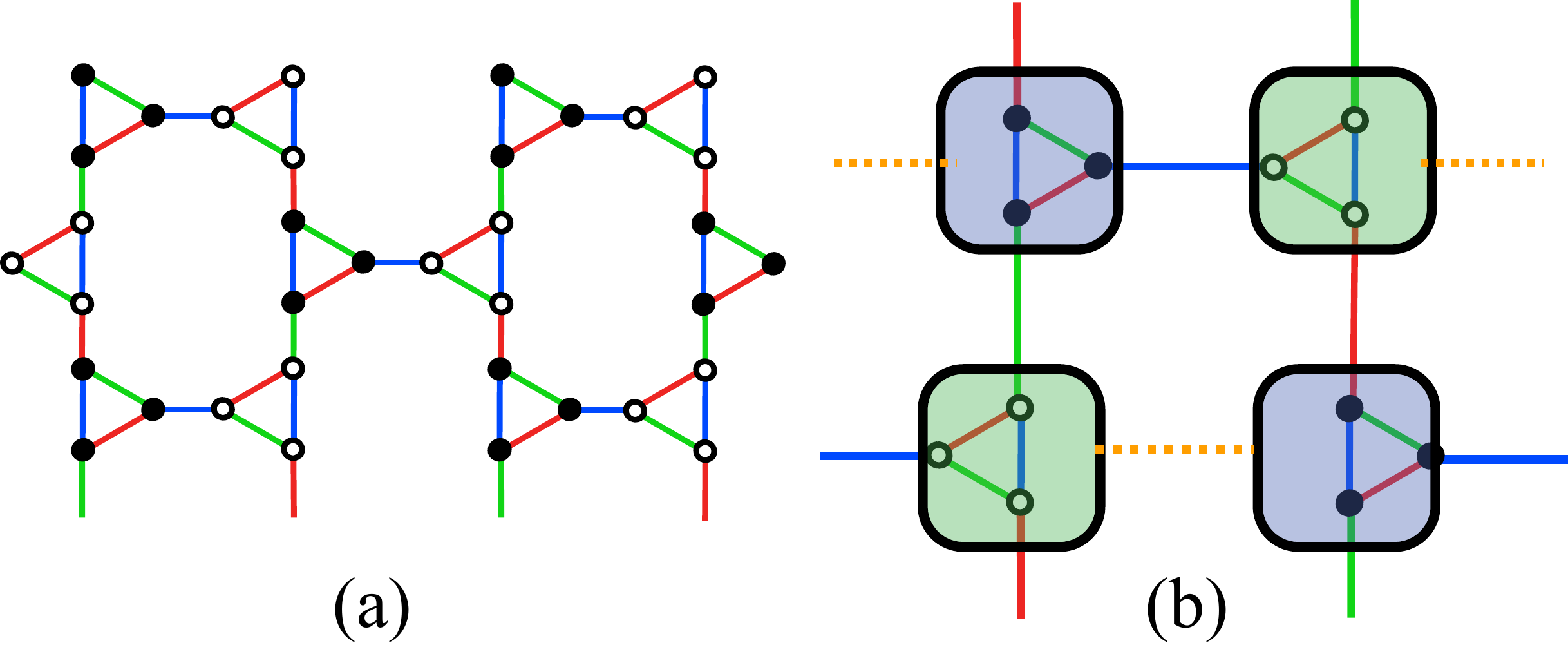}}
\caption{(Color online) (a) Star lattice with brick-wall structure. (b) Tensor network structure of a $2\times2$ unitcel of the square lattice with block sites. Each triangle of the star lattice form a block site with physical dimension $d=2^3$ for spin-$\frac{1}{2}$ models.Yellow dotted lines represent trivial bonds with $D=1$.}
\label{Fig:starblock}
\end{figure}

In this section, we briefly describe the tensor network implementation of the star lattice for general spin models in the framework of square lattice iPEPS. 

Similarly to the ruby lattice, the star lattice can be reshaped to the brick wall structure (see Fig.~\ref{Fig:starblock}-(a)) which is topologically equivalent to the star lattice represented in Fig.~\ref{Fig:ruby}-(b). Then by replacing each triangle of the star lattice with a block site with physical Hilbert space $d^3$, where $d$ is local basis of a single site, we end up with a honeycomb lattice of block sites. Associating a tensor to each block site and linking the empty edges with trivial bond dimension $D=1$, the tensor network structure of the system on the square lattice is obtained. Fig.~\ref{Fig:starblock}-(b) illustrates a $2\times2$ unit cell of the square lattice with block sites for general models on the star lattice and independent of the underlying Hamiltonian. The iPEPS implementation of the model Hamiltonian on the star lattice is problem dependent and is quite similar to the procedure described in Appendix.~\ref{appx:KR-IPEPS}.

\bibliography{references}{}
\bibliographystyle{apsrev4-1}

 \end{document}